\documentclass[twocolumn]{aastex62}

\usepackage{amsmath,amssymb,amstext}
\usepackage{mathtools}
\usepackage{gensymb}
\usepackage{graphicx}
\usepackage{scrextend}
\usepackage{natbib}
\usepackage{multirow}
\usepackage{soul}
\usepackage{chngcntr}

\newcommand{\rp}{\emph{r}-process}

\newcommand{\ye}{\ensuremath{Y_{\rm e}}}
\newcommand{\msun}{\ensuremath{M_\odot}}
\newcommand{\mej}{\ensuremath{M_{\rm ej}}}
\newcommand{\vej}{\ensuremath{v_{\rm ej}}}
\newcommand{\bt}{\ensuremath{\beta}}
\newcommand{\al}{\ensuremath{\alpha}}
\newcommand{\gm}{\ensuremath{\gamma}}
\newcommand{\edoti}{\ensuremath{\dot{\epsilon}_{\rm i}}}
\newcommand{\iso}[2]{\ensuremath{^{#2}}#1}
\newcommand{\kav}{\ensuremath{\langle \kappa \rangle}}
\newcommand{\eeff}[1]{\ensuremath{\dot{\epsilon}_{\rm eff{#1}}}}
\newcommand{\xlan}[1]{\ensuremath{X^{#1}_{\rm L,A}}}
\newcommand{\kross}{\ensuremath{\bar{\kappa}_{\rm R}}}

\newcommand{\chilan}{\ensuremath{\chi_{\rm ion,lanth.}}}

\newcommand{\dd}[1]{\ensuremath{\mathrm{d}#1}}
\newcommand{\ki}{\ensuremath{\kappa_{\rm i}}}
\newcommand{\kf}{\ensuremath{\kappa_{\rm f}}}
\newcommand{\eint}{\ensuremath{E_{\rm int}}}
\newcommand{\tlc}{\ensuremath{t_{\rm lc}}}
\newcommand{\ddf}[2]{\ensuremath{\frac{\mathrm{d}#1}{\mathrm{d}#2}}}
\newcommand{\tk}{\ensuremath{t_{\kappa}}}
\newcommand{\tauk}{\ensuremath{\tau_\kappa}}
\newcommand{\tauq}{\ensuremath{\tau_{\rm Q}}}
\newcommand{\qdot}{\ensuremath{\dot{Q}}}

\newcommand{\mone}{\textsf{frdm.y16}}
\newcommand{\mtwo}{\textsf{frdm.y28}}

\newcommand{\mthree}{\textsf{hfb22.y16}}
\newcommand{\mfour}{\textsf{hfb27.y16}}
\newcommand{\mfive}{\textsf{dz33.y16}}
\newcommand{\msix}{\textsf{unedf.kz.y16}}
\newcommand{\msev}{\textsf{unedf.xr.y16}}
\newcommand{\meig}{\textsf{unedf.y24}}
\newcommand{\mnine}{\textsf{sly4.y18}}
\newcommand{\mten}{\textsf{sly4.y21}}
\newcommand{\melev}{\textsf{tf.y16}}

\defcitealias{Barnes_etal_2016}{B16}
\defcitealias{Zhu.ea_kntherm_in.prep}{Z20}

\shorttitle{Kilonova Light-Curve Uncertainties} 
\begin{document}

\title{Kilonovae across the nuclear physics landscape: The impact of nuclear physics uncertainties on \emph{r}-process-powered emission}

\author[0000-0003-3340-4784]{Jennifer Barnes}
\affiliation{ Department of Physics and Columbia Astrophysics Laboratory, Columbia University, NY 10027 USA}

\author[0000-0003-0245-827X]{Y.~L. Zhu}
\affiliation{Department of Physics, North Carolina State University, Raleigh, NC 27695 USA}
\affiliation{Joint Institute for Nuclear Astrophysics - Center for the Evolution of the Elements, USA}

\author[0000-0003-0031-1397]{K.~A. Lund}
\affiliation{Department of Physics, North Carolina State University, Raleigh, NC 27695 USA}
\affiliation{Theoretical Division, Los Alamos National Laboratory, Los Alamos, NM, 87545, USA}

\author[0000-0002-4375-4369]{T.~M. Sprouse}
\affiliation{Theoretical Division, Los Alamos National Laboratory, Los Alamos, NM, 87545, USA}

\author[0000-0002-3305-4326]{N. Vassh}
\affiliation{University of Notre Dame, Notre Dame, Indiana 46556, USA}

\author[0000-0001-6811-6657]{G.~C. McLaughlin}
\affiliation{Department of Physics, North Carolina State University, Raleigh, NC 27695 USA}
\affiliation{Joint Institute for Nuclear Astrophysics - Center for the Evolution of the Elements, USA}

\author[0000-0002-9950-9688]{M.~R. Mumpower}
\affiliation{Theoretical Division, Los Alamos National Laboratory, Los Alamos, NM, 87545, USA}
\affiliation{Center for Theoretical Astrophysics, Los Alamos National Laboratory, Los Alamos, NM, 87545, USA}
\affiliation{Joint Institute for Nuclear Astrophysics - Center for the Evolution of the Elements, USA}

\author[0000-0002-4729-8823]{R. Surman}
\affiliation{University of Notre Dame, Notre Dame, Indiana 46556, USA}
\affiliation{Joint Institute for Nuclear Astrophysics - Center for the Evolution of the Elements, USA}

\email{jlb2331@columbia.edu}

\begin{abstract}
Merging neutron stars produce ``kilonovae''---electromagnetic transients powered by the decay of unstable nuclei synthesized via rapid neutron capture (the \emph{r}-process) in material that is gravitationally unbound during inspiral and coalescence. 
Kilonova emission, if accurately interpreted, can be used to characterize the masses and compositions of merger-driven outflows, helping to resolve a long-standing debate about the origins of \emph{r}-process material in the Universe.
We explore how the uncertain properties of nuclei involved in the \emph{r}-process complicate the inference of outflow properties from kilonova observations. 
Using \emph{r}-process simulations, we show how nuclear physics uncertainties impact predictions of radioactive heating and element synthesis.
For a set of models that span a large range in both predicted heating and final abundances, we carry out detailed numerical calculations of decay product thermalization and radiation transport in a kilonova ejecta with a fixed mass and density profile.
The light curves associated with our models exhibit great diversity in their luminosities, with peak brightness varying by more than an order of magnitude.
We also find variability in the shape of the kilonova light curves and their color, which in some cases runs counter to the expectation that increasing levels of lanthanide and/or actinide enrichment will be correlated with longer, dimmer, redder emission.
\end{abstract}


\section{Introduction} \label{sec:intro}

Electromagnetic (EM) follow-up of gravitational waves (GWs) from the binary neutron star (NS) merger GW170817 \citep{Abbott.ea_2017ApJL_gw.170817.multimess} revealed a multi-faceted EM counterpart shining at wavelengths from \gm-rays to radio waves \citep{Alexander.ea_2017.ApJ_gw.170817.radio.jet,Aracavi.ea_2017Natur_gw170817.lco.emcp.disc,Chornock.ea_2017ApJ_gw.170817.em.red.spec,Drout.ea_2017Sci_gw.170817.emcp.disc,Evans.ea_2017Sci_gw.170817.em.blue.spec, Goldstein.ea_2017ApJ:_grb.170817a.FermiGBM, Kasliwal.ea_2017Sci_gw.170817.em.interp,Kilpatrick.ea_2017.Sci_gw.170817.spectrum.opt.nir,Margutti.ea_2017.ApJ_gw.170817.xrays,McCully.ea_2017ApJ_gw.170817.blue.spec, Nicholl.ea_2017ApJ_gw.170817.blue.spec, Savchenko.ea_2017ApJ_grb.170817A.Integral, Shappee.ea_2017.Science_kn170817,Smartt.ea_2017Natur_gw170817.empc.disc,SoaresSantos.ea_2017ApJ_gw.170817.empc.decam.disc,Troja.ea_2017Natur_gw.170817.emcp.xray}.

The quasi-thermal emission observed in optical and near infrared (NIR) bands was determined by many groups \citep[e.g.][]{Drout.ea_2017Sci_gw.170817.emcp.disc,Kasen.ea_2017Natur_gw.170817.knova.theory,Kasliwal.ea_2017Sci_gw.170817.em.interp,Murgui-Berthier.ea_2017.ApJ_gw.170817.bns.interp,Tanaka.ea_2017PASJ_gw.170817.knova.interp,Waxman.ea_2018MNRAS_gw.170817.knova.interp} to conform to theoretical predictions \citep{Li_Paz_1998,Metzger_2010,Roberts_2011,Barnes_2013,Grossman_2014_kNe,Wollaeger.ea_2018MNRAS_knova.lc.morph,Fontes.ea_2020MNRAS_rporc.opacs.line.smear} of ``kilonovae," radioactive transients powered by the decay of  nuclei assembled via rapid neutron capture \citep[the \rp;][]{Lattimer_1974,Lattimer_1976,Symbalisty.Schramm.1982_rporc.nsmerg} in outflows ejected from merging neutron stars.

Atomic physics arguments \citep{Kasen_2013_AS,Tanaka_Hotok_rpOps} suggested that ejecta enriched with certain heavy \rp{} elements---lanthanides and actinides---would have a uniquely high opacity, and as a result would produce a transient whose emission peaked in the NIR.
Outflows that underwent a ``light'' \rp{}, and failed to burn these heavy elements, would instead shine in blue and optical bands \citep{Metzger_2010,Roberts_2011,Metzger.Fernandez_2014_red.blue.bhform.knova}.
Observations of the GW170817 kilonova implied two (or more) components with distinct patterns of nucleosynthesis \citep{Cowperthwaite.ea_2017ApJ_gw.170817.emcp.lcs.models,Kasen.ea_2017Natur_gw.170817.knova.theory,Perego.ea_2017ApJ_gw.170817.knova.model.3comp,Villar.ea_2017ApJ_gw.170817.knova.agg}.
The surprisingly red color of one of the components pointed to the production of at least some lanthanides, and thus established outflows from NS mergers as sites of \rp{} nucleosynthesis. 

Merger-driven mass ejection and nucleosynthesis overlap with a range of open questions in astrophysics.
These include the equation of state of NSs, which determines the ability of merging binaries to launch outflows \citep{Oechslin_2007,Bauswein_2013,Hotokezaka_2013_massEj,Kyutoku.ea_2013_mass.ej.nsbh.asym,Sekiguchi.Kiuchi.ea_2015_mass.ej.hi.ye,Lehner.ea_2016_ns.merg.asymb.binary,Sekiguchi.ea_2016.PhRvD_nms.mass.ej.asym,Dietrich.ea_2017_nsm.spin.mej,Margalit.Metzger_2017ApJ_gw.170817.ns.Mmax}; and the sites of \rp{} production throughout time and space, including whether events other than NS mergers are required to explain the heavy element enrichment of the Universe \citep{Qian_2000.ApJ_rpoc.nsm.v.sne,Argast.ea_2004.A&A_rproc.ccsne.rproc,McLaughlin.Surman_2004.NucPhysA_rproc.grb.disk,Fryer.ea_2006.ApJ_rproc.sn.fallback,Surman.McLaughlin.ea_2008.ApJL_rpoc.accdisk.nsbh,Banerjee.ea_2011.PRL_rproc.he.shells,Winteler.ea_2012.ApJ_mhd.sne.rproc.early,Moesta.ea_2018ApJ_mhdsn.rproc,Cote.ea_2018ApJ_rproc.sources.mw,Cote.ea_2019ApJ_rproc.other.sources,Siegel.ea_2019Natur_collaps.rproc,Ji.Drout.Hansen_2019.ApJ_rproc.xlan.stars.nsm,Holmbeck.Frebel.ea_2020arXiv_mass.nsm.rpoc.stars}.

Progress on these questions hinges on the ability to accurately infer ejected mass from observations of kilonovae.
The kilonova mass-luminosity relationship depends on the rate at which energy is released by radioactive decay (the ``absolute'' heating rate) and the efficiency with which that energy is converted into the thermal photons that constitute the radioactive transient.
Though thermalization efficiency has been explored before, earlier work focused on the effects of ejecta parameters like mass and velocity \citep{Barnes_etal_2016}, or employed a limited or simplified description of \rp\ radioactivity \citep{Hotokezaka_2015_heating,Kasen.Barnes_2019_analytic.therm.knova,Waxman.ea_2019_knova.late.therm,Hotokezaka.Nakar_2020ApJ_radio.heat.therm}.
Here, we explore thermalization in the context of nuclear physics uncertainties that affect predictions of the synthesis and decay of \rp{} nuclei.

Given the sheer number of nuclei that participate in the \rp{}, and the lack of experimental measurements for most of them, simulations of the \rp{} are sensitive to the 
models used to generate the required nuclear data.
This uncertainty is compounded by the dependence of the \rp{} on conditions in the outflowing gas from the time neutron capture begins.
Varying parameters in either of these categories affects the evolution of the \rp ing composition and the final abundances synthesized.
More significantly for thermalization,  it also impacts \rp{} decay, influencing the absolute heating rate, and the distribution of decay energy among decay modes and parent nuclei. 
Variability in \rp{} nucleosynthesis and decay is well-documented \citep[e.g][]{Qian.Woosley_1996_rproc.param.analytic,Beun:2007wf,Malkus:2012ts,Wanajo.ea_2014.ApJ_rproc.dyn.ej.nsm,Lippuner.Roberts_2015_rproc.params.knova,Mumpower:2015ova}.
However, while the implications of this variability for heating and thermalization have been recognized \citep{Barnes_etal_2016,Rosswog.ea_2017CQuGrav_knova.detect.adecay,Zhu.ea.Mumpower_2018_Cf254.knova,Kasliwal.ea_2019_spitzer.heavy.elems,Wu.Barnes.ea_2019_late.time.knova.rproc},
they have not before been studied systematically using detailed numerical simulations.

Here, we conduct a survey of \rp{} variability (\S\ref{sec:nucphys}), exploring how the nuclear mass model, the treatment of fission, and the initial gas conditions affect \rp{} final abundances and absolute heating rates.  
We focus on cases with robust heavy element production (i.e, on red kilonova components) for two reasons. 
First, the more extreme conditions required for the heavy (v. light) \rp{} \citep[e.g.][]{Hoffman.ea_1997_rproc.params,Freiburghaus.ea_1999_proc.adiab.exp,Lippuner.Roberts_2015_rproc.params.knova} make the question of lanthanide and actinide synthesis particularly compelling.
Second, robust \rp{} production is expected to accompany NS-black hole (BH) mergers \citep[at least in cases where the binary parameters allow tidal disruption outside the innermost stable circular orbit;][]{Meyer_1989_rproce.NSdecom.cold,East.ea_2012_eccentric.bhns.mergers,Kyutoku_2015_massEj}.
NSBH mergers have not yet been observed electromagnetically, and a full audit of the uncertainties their nucleosynthesis is subject to may allow a more nuanced interpretation of their emission once they begin to be detected.

We select from a large parameter space a set of models whose properties represent the diversity allowed by astrophysical variation and nuclear physics uncertainties.
We then simulate the emission and thermalization of radioactive decay products in the kilonova ejecta with a level of detail beyond what has so far been considered in the literature, using energy-loss cross sections and particle emission rates and spectra consistent with each model's composition (\S\ref{sec:therm_results}).
We find a fair degree of variation among models, with the most important determinants of thermalization being the slope of the absolute heating curve and the energies of the emitted decay products.

These simulations determine the total ``effective'' heating rate for each model. which we use in radiation transport calculations of kilonovae (\S\ref{sec:kilonova}).
From the results, we construct synthetic bolometric light curves.
Not surprisingly, the large range in effective heating propagates through to the light curves, which exhibit considerable diversity, both in the peak brightness and the light-curve shape.
We find that varying the effective heating rate and the composition self-consistently leads, in certain cases, to unexpected outcomes, such as a heavy element-rich outflow with a surprisingly early peak.
We also discuss the kilonova spectral energy distributions, which display less variability than the bolometric light curves.
Our results argue for the careful consideration of nuclear physics assumptions in the construction of kilonova models and the interpretation of kilonova emission.

\section{Nuclear Physics}\label{sec:nucphys}

We use nuclear physics simulations to determine the potential variability in \rp\ nucleosynthesis and decay energetics, and to extract for different choices of parameters the data needed to self-consistently calculate decay-energy thermalization and kilonova EM emission.

\subsection{Nucleosynthesis Calculations} \label{subsec:nucsims}
Much of the uncertainty in simulations of \rp\ nucleosynthesis is due to the discrepant predictions of various theoretical models used to compute unmeasured properties of the nuclei involved.
While a full discussion of the nuclear physics inputs we consider is provided in \citet[][henceforth Z20]{Zhu.ea_kntherm_in.prep}, the main points are summarized here as well.

We base our theoretical nuclear inputs on each of eight distinct nuclear models listed in
Table~\ref{tab:nucparams}.
Unmeasured reaction rates, as well as \al- and \bt-decay lifetimes and branching ratios, are adjusted to take into account $Q$-values corresponding to 
each mass model in a manner consistent with the predicted nuclear masses \citepalias[][and references therein, especially \citet{Mumpower.ea_2015.PhRvC_num.masses.rproc.Y}]{Zhu.ea_kntherm_in.prep}.

The potential energy landscape of fission influences fission half-lives and the mass distribution of the daughter fragments, both of which can be important, particularly in the case of strong fission cycling.  
Our nucleosynthesis calculations include neutron-induced, \bt-delayed, and spontaneous fission, and so
the treatment of fission adds an additional dimension of uncertainty. 

Fission barrier heights have been calculated within the framework of several nuclear models, but have not been calculated for every model we consider here. 
When possible, we adopt barrier heights computed for a mass model that closely resembles our model of interest. 
In cases where there is no clearly associated mass model for which barrier heights have been determined, we use the barrier heights predicted by the Finite Range Liquid Droplet Model \citep[FRLDM;][]{Moller.ea_2015PRC_frldm.fiss.barrs}. 

Our \rp{} simulations use phenomenological descriptions of the fission fragment mass distribution that do not take barrier heights as input.
The distribution is taken to be either Gaussian, according to the formalism of \citet{Kodama.Takahashi.ea_1975NPA_rproc.fission}, or symmetric ($A_{\rm daught} = A_{\rm parent}/2$). 
The exception is \iso{Cf}{254}, for which we employ fission yields calculated with FRLDM \citep{Mumpower2020}, as in \citet{Zhu.ea.Mumpower_2018_Cf254.knova}. 
We also use phenomenological prescriptions for spontaneous fission lifetimes, calculating these rates using the either the method of \citet{Xu.Ren_2005PR_spont.fiss.thalf} or the barrier height-dependent prescription of \citet{Karpov.ea_2012ijmpe_heavy.elem.fiss} and \citet{Zagrebaev.ea_2011PRC_nrich.heavy.fiss}.

Since spontaneous fission plays an important role in heating, the four combinations of fission yields and spontaneous fission rates are intended to bracket the range of possible fission behavior. 
The rates of  \citet{Xu.Ren_2005PR_spont.fiss.thalf} become substantial around $Z>94$, while those of \citet{Zagrebaev.ea_2011PRC_nrich.heavy.fiss} which depend explicitly on fissibility $(Z^2/A)$ and barrier height, are typically fairly low until $Z > 100$. \citep[For a more detailed discussion of these two spontaneous fission models, see][]{Vassh:2020bzs}.  
Meanwhile, a symmetric split  produces a very narrow set of daughter products, while the yields of \citet{Kodama.Takahashi.ea_1975NPA_rproc.fission} are more broadly distributed. 

In addition to nuclear physics uncertainties, the \rp\ is also sensitive to the outflow properties of the \rp ing gas. 
While the thermodynamic and hydrodynamic properties of the outflow (entropy and expansion rate, e.g.) can impact the \rp, in the regime of robust heavy element production (i.e., in our regime of interest), nucleosynthesis has been shown to be particularly sensitive to the initial electron fraction \ye\ \citep{Freiburghaus_1999,Goriely_2011,Lippuner.Roberts_2015_rproc.params.knova}, defined as the ratio of protons to baryons in the gas.
We therefore use \ye\ to probe the dependence of the \rp\ on astrophysical conditions, and consider for each set of nuclear physics inputs eight values in the range $0.02 \leq \ye \leq 0.28$.
The nuclear physics models and astrophysical quantities that define our parameter space are presented in Table~\ref{tab:nucparams}. 
All possible combinations of these parameters are considered, yielding a full set of 256 models. 
\begin{table}
\bigskip
\centering
\caption{Nucleosynthesis Parameter Space for the Full Set of Simulations}\label{tab:nucparams}
\begin{tabular}{>{\raggedright\arraybackslash}p{0.45\columnwidth} >{\raggedleft\arraybackslash}p{0.45\columnwidth}}
\toprule
\multicolumn{2}{c}{\rule{0pt}{2.4ex} \bf Nuclear Mass Models }\\
\rule{0pt}{2.4ex}\textit{Mass Model} & \textit{Reference} \\
\hline
\rule{0pt}{2.6ex}Finite Range Droplet Model (FRDM2012)  & \multirow{2}{*}{\citet{Moller.ea_2016ADNDT_frdm2012}} \\
\rule{0pt}{2.7ex}Duflo \& Zuker (DZ33) & \citet{Duflo.Zuker_1995PRC_DZ.mass.model} \\
\rule{0pt}{2.7ex}Hartree-Fock-Bogoliubov 22 (HFB22) & \multirow{2}{*}{\citet{Goriely.ea_2013PRC_HFB.mass.model.27}} \\
\rule{0pt}{2.7ex}Hartree-Fock-Bogoliubov 27 (HFB27) & \multirow{2}{*}{\citet{Goriely.ea_2013PRC_HFB.mass.model.27}} \\
\rule{0pt}{2.7ex}Skyrme-HFB with UNEDF1 (UNEDF1) & \multirow{2}{*}{\citet{Kortelainen.ea_2012PRC_unedf1.mass.model}} \\
\rule{0pt}{2.7ex}Skyrme-HFB with SLY4 (SLY4) & \multirow{2}{*}{\citet{Chabanat.ea_1998NPA_skyrme.hfb}} \\
\rule{0pt}{2.7ex}Extended Thomas-Fermi plus Strutinsky Integral (ETFSI) &  \multirow{2}{*}{\citet{Aboussir.ea_1995ADNDT_etsfi.mass.model}} \\
\rule{0pt}{2.7ex}Weiz\"{a}cher-Skyrme (WS3)& \citet{Liu.ea_2011PRC_nuc.masses.ws3}\rule[-1.2ex]{0pt}{0pt}\\
\end{tabular}
\begin{tabular}{>{\raggedright\arraybackslash}m{0.25\columnwidth} >{\raggedleft\arraybackslash}p{0.65\columnwidth}}
\toprule
\multicolumn{2}{c}{\rule{0pt}{3ex}\bf Fission Prescription  }\\
\rule{0pt}{2.4ex}\textit{Mass Model}  & \it Barrier Heights Adopted\\
\hline 
\rule{0pt}{2.6ex}HFB22 & HFB14 \citep{Goriely.ea_2009.PhRvC_HFB.fiss.barriers}\\
HFB27 & HFB14 \\
ETFSI & ETFSI \citep{Mamdouh.ea_2001.NucPhysA_etfsi.fiss.barriers}\\
\multirow{2}{*}{All others} & Finite-Range Liquid-Droplet Model \\
& \citep[FRLDM;][]{ Moller.ea_2015PRC_frldm.fiss.barrs} \\
\end{tabular}
\begin{tabular}{>{\raggedright\arraybackslash}p{0.45\columnwidth} >{\raggedleft\arraybackslash}p{.45\columnwidth}}
\rule{0pt}{3.6ex}\it Fission Fragment & \multirow{2}{*}{\it Reference}\\
\it Distribution \\
\hline
\rule{0pt}{2.6ex}Symmetric & e.g.,  \citet{Rauscher.ea_1994ApJ_big.bang.rporc.fiss.sym} \\
Gaussian  & \citet{Kodama.Takahashi.ea_1975NPA_rproc.fission} \\

\rule{0pt}{3.6ex}\it Spontaneous & \multirow{2}{*}{\it Reference} \\
\it Fission Lifetimes \\
\hline
\rule{0pt}{2.6ex}Xu \& Ren (XuRen) & \citet{Xu.Ren_2005PR_spont.fiss.thalf} \\
\multirow{2}{*}{Karpov/Zagrebaev (KZ)}  & \citet{Karpov.ea_2012ijmpe_heavy.elem.fiss, Zagrebaev.ea_2011PRC_nrich.heavy.fiss}\rule[-1.2ex]{0pt}{0pt} \\

\end{tabular}
\begin{tabular}{>{\raggedright\arraybackslash}m{0.1\columnwidth} >{\raggedleft\arraybackslash}p{0.8\columnwidth}}
\toprule
\multicolumn{2}{c}{\rule{0pt}{3ex}\bf Astrophysical Parameters  }\\
\rule{0pt}{2.4ex}\ye  & 0.02, 0.12, 0.14, 0.16, 0.18, 0.21, 0.24, 0.28 \\
$s_{\rm B}$ & $40 k_{\rm B}$ \\
$\tau_{\rm exp}$ & 20 ms\rule[-1.2ex]{0pt}{0pt} \\
\hline
\end{tabular}
\end{table}

We use the Portable Routines for Integrated nucleoSynthesis Modeling
\citep[PRISM;][]{Sprouse.ea_PRISM_in.prep}
to simulate the \rp\ for each of our 256 models for a gas trajectory with an initial entropy per baryon $s_{\rm B} = 40 k_{\rm B}$ (with $k_B$ the Boltzmann constant) and an expansion timescale $\tau_{\rm exp} = 20$ ms during the \rp{} epoch.
This trajectory has also been used in other studies of post-merger nucleosynthesis \citep[][\citetalias{Zhu.ea_kntherm_in.prep}]{Zhu.ea.Mumpower_2018_Cf254.knova}.
The choices of $s_{\rm B}$ and $\tau_{\rm exp}$ are motivated by models of disk wind outflows from NS mergers, and are therefore consistent with our focus on the heavy \rp\ (red) kilonova component, which both spectral analysis \citep{Chornock.ea_2017ApJ_gw.170817.em.red.spec, Kasen.ea_2017Natur_gw.170817.knova.theory} and theoretical simulations \citep{Siegel.Metzger_2018_3d.grmhd.disk.sims} suggest is associated with a disk wind (though see \citet{Miller.ea_2019_disk.blue.knova.nsm} for an alternate view).
We set the initial temperature to 10 GK for all calculations, and compute the seed nuclear abundances with the SFHo equation of state \citep{Steiner.Hempel.Fischer_2013.apj_eos.ccsne}.
The temperature evolution of the expanding gas is determined self-consistently for each simulation using the same reheating procedure as in \citet{Zhu.ea.Mumpower_2018_Cf254.knova} and \citet{Vassh.Vogt.Surman.ea_2019.JPG_rproc.fission.yields}.

PRISM tracks the nuclear abundances for each model as a function of time.
The absolute nuclear heating is then computed from the evolving composition, using known or predicted decay modes and branching ratios for all nuclei, and measured or theoretical nuclear masses, which allow a determination of the energy $Q$ released in each decay.
The absolute heating rate, defined as the total energy emitted by radioactive decays per unit mass and time, is then a sum over decays.

\subsection{Selection of Models}\label{subsec:modprops}

The calculations performed on the full set of 256 models delineated, for the parameter space under consideration, the  potential variation in total \rp\ heating. From the full set of parameter combinations, we selected 10 models whose absolute nuclear heating spans the range indicated by the full set.

\begin{table*}[t]
\begingroup
\setlength{\tabcolsep}{3.5pt}
\bigskip
\centering
\caption{Properties of the Model Suite}\label{tab:models}
\begin{tabular*}{\textwidth}{>{\centering\arraybackslash}p{0.1\textwidth}>{\centering}p{0.12\textwidth}>{\centering}p{0.14\textwidth}>{\centering}p{0.22\textwidth}>{\centering}p{.1\textwidth}>{\centering}p{0.12\textwidth}|>{\centering\arraybackslash}p{0.08\textwidth}}
\toprule
\rule{0pt}{2.4ex}\multirow{2}{*}{\textit{Model Index}}&\textit{Nuclear Mass Model} & \textit{Fission Fragment Distribution} & \textit{Spontaneous Fission\\ $\tau_{\rm nuc}$} & \multirow{2}{*}{\ye} & \multirow{2}{*}{\textit{Model Label}} & \multirow{2}{*}{$X_{\rm L,A}^\dagger$} \\
\hline
\rule{0pt}{2.6ex}1 & FRDM2012 & Symmetric & Karpov/Zagrebaev & 0.16 &  frdm.y16 & 0.22 \\
2 & FRDM2012 & Symmetric & Karpov/Zagrebaev & 0.28 &  frdm.y28 & 0.02 \\
3 & HFB22 & Symmetric &Karpov/Zagrebaev & 0.16  & hfb22.y16 & 0.37\\
4 & HFB27 & Gaussian  & Karpov/Zagrebaev & 0.16 & hfb27.y16  & 0.48 \\
5 & DZ33 & Gaussian & Karpov/Zagrebaev & 0.16 & dz33.y16 & 0.31 \\
6 & UNEDF1 & Gaussian & Karpov/Zagrebaev & 0.16 & unedf.kz.y16 & 0.34\\
7 & UNEDF1 & Gaussian & Xu \& Ren & 0.16 & unedf.xr.y16 & 0.33\\
8 & UNEDF1 & Symmetric & Karpov/Zagrebaev & 0.24 & unedf.y24 & 0.38\\
9 & SLY4 & Symmetric & Karpov/Zagrebaev & 0.18 & sly4.y18 & 0.03 \\
10 & SLY4 & Symmetric & Karpov/Zagrebaev & 0.21 & sly4.y21 & 0.12\\
11$^\ddagger$ & TF+D3C* & Symmetric & Karpov/Zagrebaev & 0.16 & tf.y16 & 0.22 \\
\hline
\end{tabular*}
\endgroup
\hspace{0.5mm}
\footnotesize{$^\dagger$\xlan{} is a result of, not an input parameter to, our nuclear physics simulations.}
\newline \footnotesize{$^\ddagger$Model 11 was not selected from the full suite of simulations. It was instead constructed specifically to expedite comparison between this work and that of other groups. See  \S \ref{subsec:modprops} for further discussion.}
\end{table*}

By design, these models represent the maximum variation in absolute heating predicted by the full set, and some therefore appear as outliers.
However, even the full set does not account for all possible nuclear physics uncertainties, and moreover reflects our decision to focus on a limited set of mass models and fission descriptions.
The true allowed spread in heating is therefore expected to be larger even than our model set indicates.

The parameters that define our ten-model subset are recorded in Table~\ref{tab:models}.
To this set of 10, we add a single model calculated using masses and fission barriers determined by the Thomas-Fermi (TF) mass model \citep{Myers.Swiatecki_NucPhysA.1996_thomas.fermi.mmodel,Myers.Swiatecki_1999.PhRvC_TF.barrier.heights}.
TF barrier heights were calculated using the prescription applied in
the GEF code \citep[2016 version;][]{Schmidt.ea_NDS.2016_GEF.code}, as in \citet{Vassh.Vogt.Surman.ea_2019.JPG_rproc.fission.yields}. 
We include this model to explore the impact of TF barriers, which tend to be lower than the other barriers we consider in key regions in a way that limits the production of long-lived fissioning isotopes \citep{Vassh.Vogt.Surman.ea_2019.JPG_rproc.fission.yields}.
For this model, we apply the ``D3C*'' \bt-decay rates of \citet{Marketin.Huther.MPinedo_2016.PhysRevC_bdecay.rates.d3cstar}, which have been used in conjunction with Thomas-Fermi barriers for some fission channels in studies of the \rp{} \citep[e.g.][]{Wu.Barnes.ea_2019_late.time.knova.rproc}.
While we do not incorporate TF+D3C* into the nuclear physics parameter space defined in Table~\ref{tab:nucparams}, we adopt it in Model 11 to facilitate comparisons between our work and that of other groups.

\begin{figure*}\includegraphics[width=\textwidth]{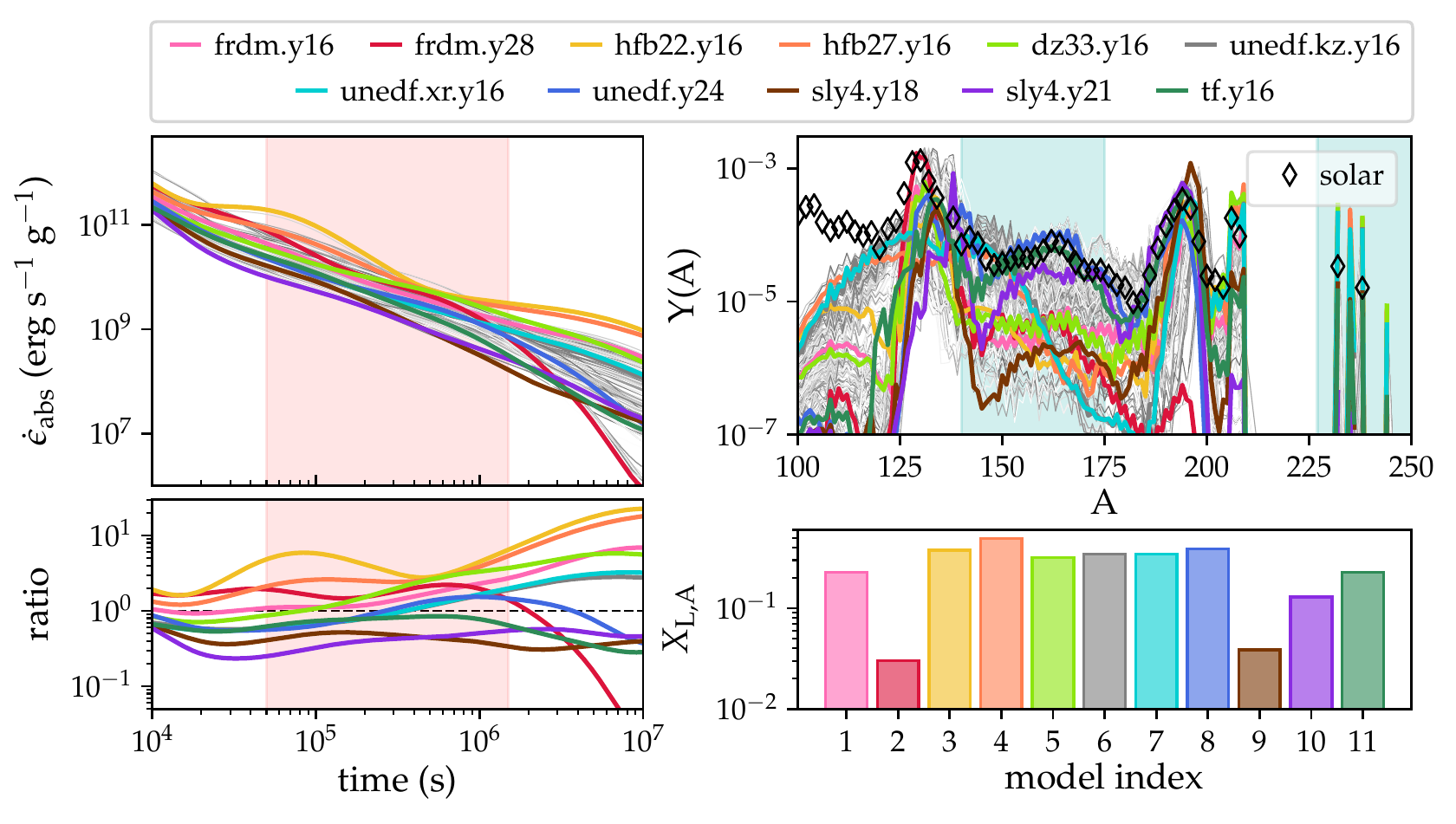}
\caption{The properties of the subset of models selected for further study.
\textit{Top left:} The absolute heating rates from radioactivity as a function of time. The heating rates vary by ${\sim}1$ order of magnitude on kilonova timescales (${\sim}0.5 - 15$ days; indicated by the transparent red bar). The heating rates for the full set of models are plotted as thin gray lines to demonstrate that the subset is representative. 
\textit{Bottom left:} The ratio of the model heating rates to the power-law heating rate $\dot{E}_{\rm abs} = 2 \times 10^{10} (t/{\rm day})^{-1.3}$, typical of analytic approximations to \rp\ heating. The approximation captures only the most basic behavior of the numerically determined heating rates, which deviate from a straightforward power law, at times considerably.
\textit{Top right:} The final abundance patterns of our models. Scaled solar abundances from \citet{Sneden.Cowan.Gallino_araa.2008_n.capt.elems} are plotted as black diamonds. (For $A<220$, we plot data only for even $A$ to preserve readability.) The final abundance patterns of the full model set are plotted as thin gray lines, and the transparent teal bars show the rough positions of the high-opacity lanthanides and actinides. \textit{Bottom right:} The total lanthanide and actinide mass fraction of each model at $t=1$ day.}
\label{fig:model_qandy}
\end{figure*}

The absolute heating rate and abundance information for each model in our eleven-model subset are presented in Figure~\ref{fig:model_qandy}.
Like the full suite of simulations (the results of which we plot in gray scale for comparison), the model subset exhibits significant diversity in both the energy produced and the composition synthesized.
The first of these has implications for thermalization, which affects the amount of energy available to power the light curve, while the latter is important for determining opacity, which is sensitive to the abundances of lanthanides and actinides in the ejecta. 

In the interest of fully exploring the potential variability in \rp\ heating, we did not restrict ourselves to models whose abundance patterns conformed to the solar \emph{r}-pattern, and the abundance yields for our models vary substantially. 
We did ensure that all our models had a combined lanthanide and actinide mass fraction, \xlan{}, high enough to plausibly produce the large opacity required to explain the red component of the G170817 kilonova.
(In our model subset, $X_{\rm L,A} \geq 0.02$ at $t \approx 1$ day; see Table~\ref{tab:models}.)

In fact, some models have \xlan{} greater than typically considered in studies of kilonovae ($\xlan{\rm typ} \lesssim 0.1$).
The effects of higher \xlan{} are explored in \S\ref{sec:kilonova}.
Here, we simply affirm that consideration of high-\xlan{} models is warranted given a) their potential relevance for NSBH kilonovae, b) indications from \rp-enriched stars that a sizeable fraction of \rp{} events should have \xlan{} greater than is usually attributed to the kilonova of GW170817 \citep{Ji.Drout.Hansen_2019.ApJ_rproc.xlan.stars.nsm}, and c) observations of an actinide ``boost'' in ${\sim}30$\% metal poor stars \citep{Mashonkina.ea_2014.A&A_actin.boost.stars}, seeming to require copious actinide production by the \rp{} \citep{Holmbeck.ea_2019.ApJ_rproc.actin.boost}.
We also note that the imperfect (though still impressive) agreement between models and observations \citep[e.g.][]{Chornock.ea_2017ApJ_gw.170817.em.red.spec,Kasliwal.ea_2017Sci_gw.170817.em.interp,Smartt.ea_2017Natur_gw170817.empc.disc} of the GW170817 kilonova---the lone event of its class---make it premature to conclude that high \xlan{} lie outside the realm of possibility even for NS mergers.

We carry out for these eleven models detailed numerical calculations of thermalization efficiencies and kilonova light curves.

\section{Thermalization Model}\label{sec:therm}

The variation in absolute heating from nuclear decay is one important source of uncertainty in kilonova models.
However, kilonova luminosity is not set solely by the energy released in radioactivity.
It is instead a function of the portion of radioactive energy converted into the thermal photons that ultimately form the light curve.
In addition to the properties of the ejecta (e.g., its mass and kinetic energy) this \emph{thermalization efficiency} depends on the rate at which radioactivity releases energy into the ejecta, the partition of that energy among different decay channels, the spectra of the emitted particles, and the cross sections for the interactions by which decay products transfer their kinetic energy to the thermal pool, which themselves depend on the ejecta composition \citep{Barnes_etal_2016, Kasen.Barnes_2019_analytic.therm.knova,Waxman.ea_2019_knova.late.therm,Hotokezaka.Nakar_2020ApJ_radio.heat.therm}.
Thermalization efficiency is therefore not universal; the same uncertainties that give rise to variability in absolute heating will also impact thermalization. 
A true accounting of the kilonova diversity allowed by nuclear and astrophysical uncertainties must include a self-consistent determination of thermalization efficiency.

We undertake detailed, numerical calculations of thermalization for each model in our subset. 
We use results of the \texttt{PRISM} simulations along with available experimental data to characterize the radioactive decay profile and the ejecta composition for our subset of models.
Then, using a modified version of the framework established in \citet[][henceforth \citetalias{Barnes_etal_2016}]{Barnes_etal_2016}, we calculate the emission and thermalization of radioactive decay products in the kilonova ejecta.

For each model, we simulate particle emission in a manner that preserves the total energy emitted ($\dot{\epsilon}_{\rm abs}$), its division among decay products, and the emission spectrum of each type of particle emitted by radioactive decay.
Emitted particles are transported through the ejecta, depositing their energy according to particle- and composition-specific, energy-dependent cross sections.
Improving on the work of \citetalias{Barnes_etal_2016}, we record energy deposition as a function of both time and space.
In the following, we outline how we determine the inputs for the thermalization calculation, and describe updates to the method of \citetalias{Barnes_etal_2016}.

\subsection{Partition of Radioactive Energy}\label{subsec:enpart}

Each set of nuclear physics models we consider predicts the decay modes and branching ratios of any nuclear decay for which these quantities have not been measured.
Together with experimental data, these predictions allow us to determine the fraction of the absolute heating rate contributed by the three decay channels important on kilonova timescales: \al-decay; \bt-decay; and spontaneous fission.
(\texttt{PRISM} also tracks \bt-delayed and neutron-induced fission, but these do not operate after very early times.)
The decay products from each channel have distinct energy scales and  cross sections for energy-loss processes, and so are treated individually in thermalization calculations.

In \al-decay (fission), the decay energy mainly accrues to the \al-particle (fission fragments), and so the kinetic energy of the emitted \al-particle (fission fragments) is taken to equal to the energy $Q$ of the decay.
In contrast, the energy from a \bt-decay is shared by a \bt-particle, a neutrino, and an arbitrary number of \gm-rays.
The \gm-rays thermalize only weakly, and neutrinos do not thermalize at all, so additional analysis to determine the division of \bt-decay energy among these species is necessary.

We determined at each time step the 50 nuclei producing the most energy through \bt-decay, and calculated from this subset the fraction of the total \bt-decay energy emerging as \bt-particles, \gm-rays, and neutrinos using \bt-decay endpoint energies, average kinetic energies, and intensities from the Nuclear Science References database \citep{NucData_References}, which we accessed through the website of the International Atomic Energy Agency.
We assumed parent nuclei decay from the nuclear ground state (though see \citet{Fujimoto.Hashimoto_2020MNRAS_knova.isomers} for a discussion of isomeric decay in the \rp).
When possible, we estimated unknown quantities (e.g., unrecorded average \bt-particle energies were assumed to be 1/3 of the corresponding \bt\ endpoint energy).
If there was insufficient data to build a reasonable model of a given decay, the nucleus in question was excluded from our analysis.
However, by the time the kilonova begins to rise ($t \sim$ few hours), the composition has already stabilized relative to earlier times, and exotic unmeasured isotopes are the exception rather than the rule.
Thus, few nuclei were omitted.

\begin{figure*}\includegraphics[width=\textwidth]{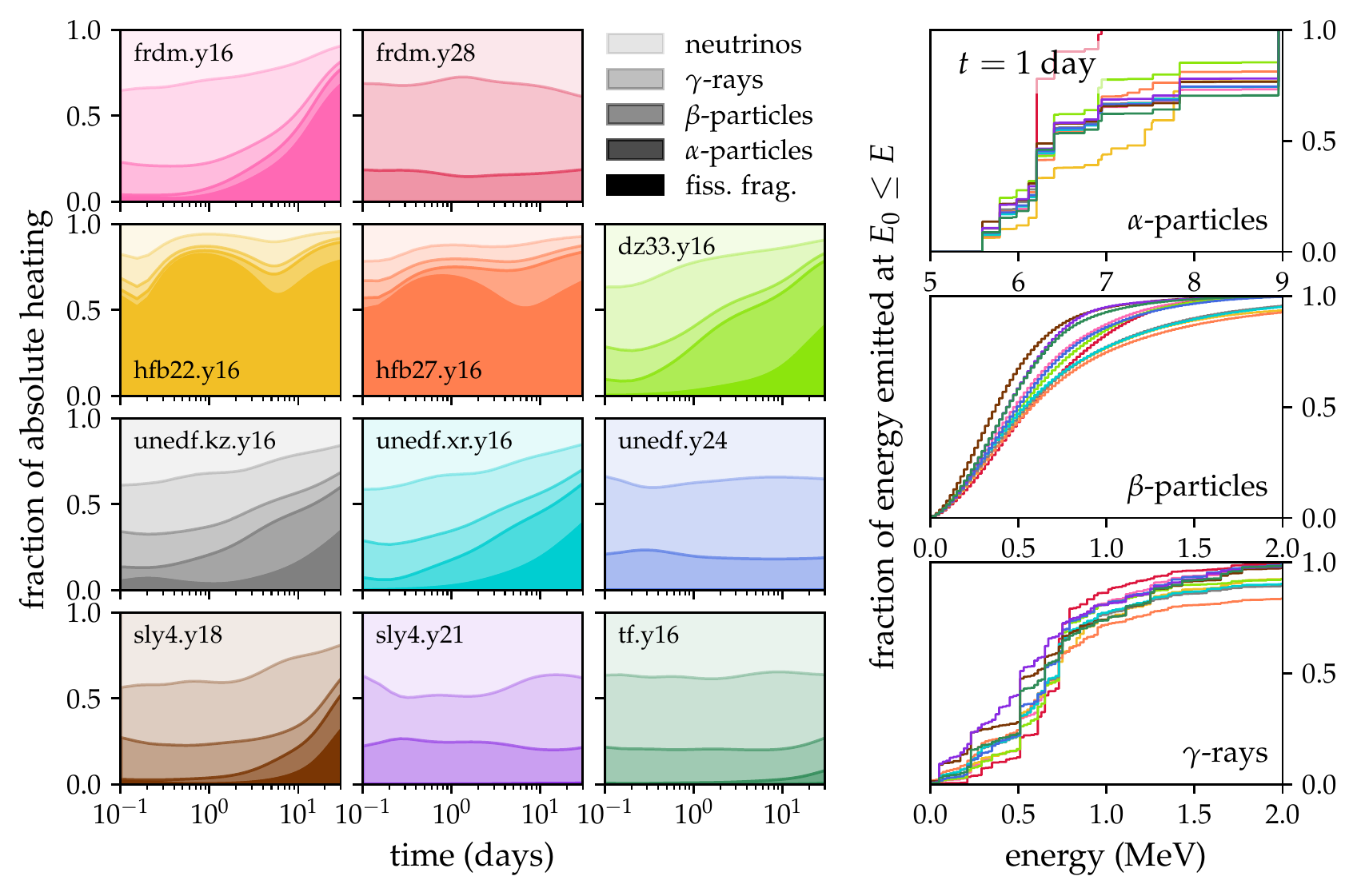}
\caption{A summary of the radioactivity profiles of our subset of models. \emph{Left panels:} The 11 panels on the left-hand side show how the absolute heating from radioactive decay is partitioned among radioactive decay products for each model. While all models have a contribution from \bt-decay, the roles of fission and \al-decay vary. \emph{Right panels:} The cumulative distribution functions of the emission spectra of \al-particles, \bt-particles, and \gm-rays for each model at $t=1$ day post-merger. The curves are color-coded using the same scheme as in the panels on the left (and as in Fig.~\ref{fig:model_qandy}). The spectra differ due to the distinct populations of decaying nuclei for each model.} 
\label{fig:radiosum}
\end{figure*}

The distribution of the absolute heating among decay products is shown for each model as a function of time in the left-hand panels of Figure~\ref{fig:radiosum}.
The models exhibit considerable diversity, particularly in regard to the importance of \al-decay and fission, which have been shown to enhance thermalization \citepalias{Barnes_etal_2016}.
Roughly half the models show the contributions from these channels growing slowly, accounting for ${\gtrsim}30\%$ of the absolute heating by $t = 30$ days. 
This increase is due to the production of certain actinide species (in the case of \al-decay) and the spontaneous fissioning isotope \iso{Cf}{254} (in the case of fission), whose long half-lives allow them to dominate the energy release at late times, once the background \bt-decay radioactivity has subsided.
Other models (\mtwo{}, \meig{}, \mten{}, and \melev{}) have only negligible contributions from \al-decay or fission, and still others (\mthree{} and \mfour{}) have heating rates that are dominated by fission over the entire kilonova timescale.

We used the particle-specific absolute heating rates extracted from the \texttt{PRISM} calculations in our thermalization simulations.
This improves on the method of \citetalias{Barnes_etal_2016}, which
assumed the absolute heating from all decay modes and particle types obeyed a power law.
Recent analytic work \citep{Kasen.Barnes_2019_analytic.therm.knova, Waxman.ea_2019_knova.late.therm} showed that the interplay between the absolute heating associated with a given decay product and that product's time-dependent emission spectrum influences thermalization, motivating the more rigorous treatment of $\dot{\epsilon}_{\rm abs}$ in this work.
The range of behavior seen in our models, in terms of both of the shape and magnitude of the absolute heating curve, and the relative importance of the different decay products to the total heating, suggests that thermalization efficiencies will be variable.

\subsection{Radioactive Emission Spectra}\label{subsec:radspec}

Because the cross sections of thermalizing processes are energy-dependent, the emission spectra of radioactive decay products are required for thermalization calculations.
We construct the time-dependent emission spectra for \al-particles, \bt-particles, and \gm-rays with a procedure similar to that of \citetalias{Barnes_etal_2016}.
We assign to emitted \al-particles the entire $Q$-value of the decay that produced them (\al-decay typically proceeds directly to the daughter ground state, so the incidence of emitted \gm-rays is vanishingly low.)
The \al-decay spectrum then simply reflects the fraction of the total \al-decay heating contributed by every \al-decaying nucleus at a given time.

The spectra of \bt-decay products are less straightforward due to the three-body nature of the decay and the fact that \bt-particle and neutrino emission generally land the daughter nucleus in one of many excited nuclear states, with de-excitation to the ground state occurring \gm-ray emission. 
We compute the \bt-particle spectrum for each nucleus from available endpoint energies and intensities, \citep{NucData_References} assuming using the simplified fit to the Fermi formula for allowed transitions proposed by \citet{Schenter_Vogel_1983_betaSpec}.
In cases where forbidden transitions are important, the true spectrum may deviate from our calculations.
The \gm-ray spectrum was determined from recorded \gm-ray energies and intensities.
The \bt-particle (\gm-ray) spectrum for the entire composition was then calculated by weighting the spectrum from each isotope by the fraction of the total energy in \bt-particles (\gm-rays) contributed by that isotope, as a function of time. 

Snapshots of the \al-particle, \bt-particle, and \gm-ray emission spectra for each model at $t = 1$ day are provided in the right-hand panels of Figure~\ref{fig:radiosum}.
The \al-spectrum is dominated by about a dozen nuclei, each of which emits an \al-particle with a single characteristic energy.
The \bt-particle and \gm-ray spectra are smoother, because of the continuous nature of the emission from individual nuclei (\bt-particles) or the much larger number of discrete emission energies (\gm-rays).

Because of the inverse relationship between $Q$ and lifetime for \bt-decaying nuclei \citep{ColgateWhite_1966_NeutrinoMech,Metzger_2010,Hotokezaka.ea_2017_rproc.edot.analytic}, \bt-decay spectra tend to shift toward lower energies as time progresses.
Even by $t=1$ day, many of the less stable (higher $Q$) isotopes have decayed away, concentrating the \bt- and \gm-ray spectra at $E \lesssim 1$ MeV.
However, late-time fission could produce a population of unstable \bt-decaying nuclei with high $Q$, in defiance of this trend \citepalias{Zhu.ea_kntherm_in.prep}.

Unlike for other decay channels, we do not attempt a full calculation of the fission fragment spectrum, which would need to be described in terms of both daughter mass $A_{\rm ff}$ and initial kinetic energy $E_{\rm ff,0}$.
However, test calculations of fission fragment thermalization that adopted $\delta$-function distributions of $A_{\rm ff}$  and $E_{\rm ff,0}$ over the ranges $100 \leq A_{\rm ff} \leq 200$ and $100 \leq E_{\rm ff,0}/\text{MeV} \leq 200$ yielded very similar results, suggesting that fission fragment thermalization is not highly sensitive to these quantities.
Therefore, though fission daughter distributions are tracked internally by PRISM to model nucleosynthesis, we do not extract from PRISM simulations a detailed description of fission fragment production for our thermalization calculations.
Rather, we adopt a fiducial fragment mass $A_{\rm ff} = 150$ and a flat emission spectrum that ranges from 100 -- 150 MeV.

\subsection{Energy Transfer by Radioactive Decay Products}

\begin{figure}\includegraphics[width=\columnwidth]{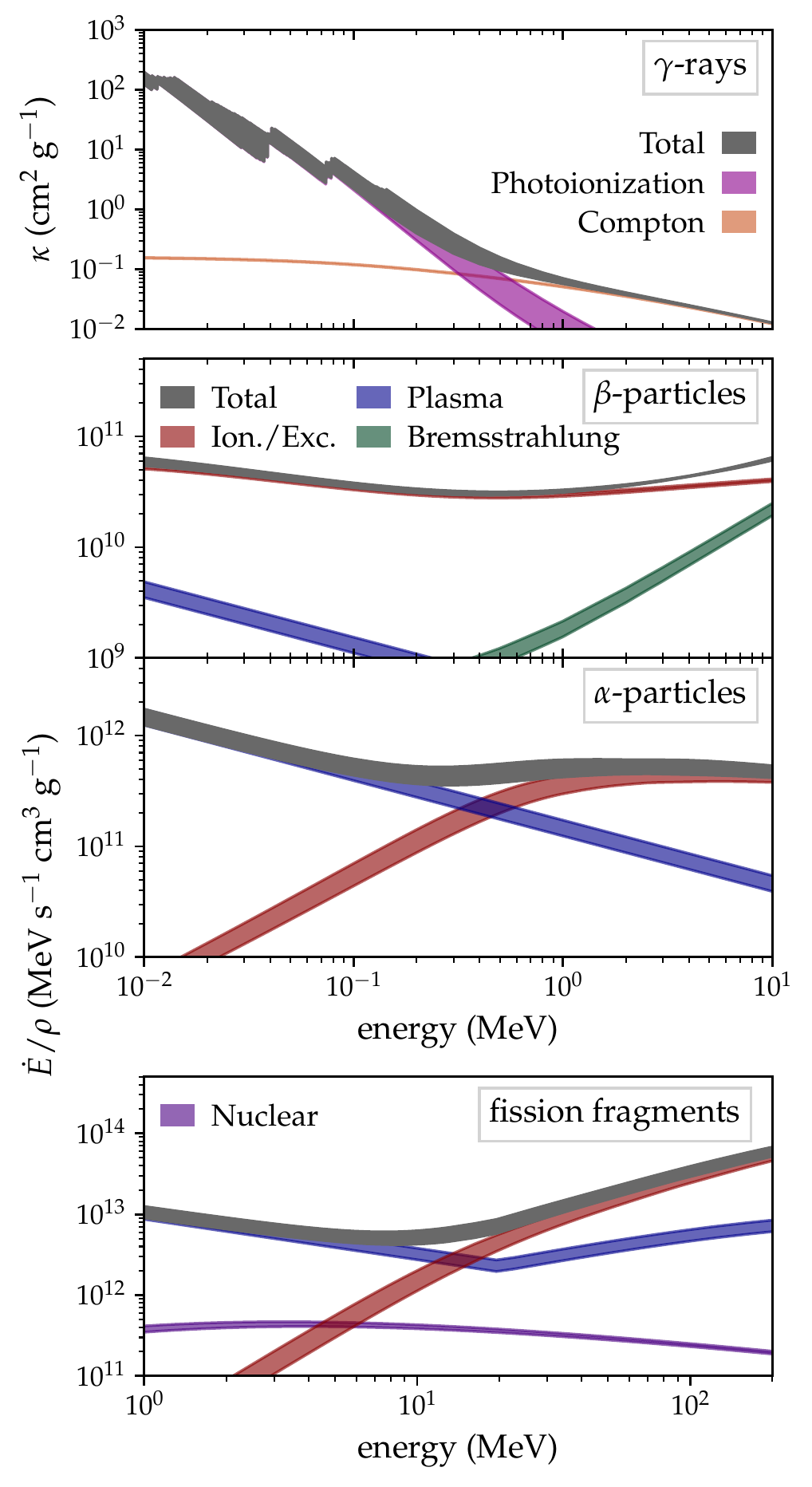}
\caption{The range of energy-loss rates and cross sections associated with our model subset.
All rates and cross sections were calculated for the model compositions as measured at $t=1$ day; the evolution of the composition beyond that point was not found to change the rates meaningfully.
Energy-loss rates (bottom three panels) have been normalized to mass density.
\emph{Top panel:} Both Compton scattering (orange band) and photoionization (fuchsia band) contribute to the \gm-ray opacity.
The former (latter) dominates at high (low) energies.
\emph{Bottom three panels:} Energy-loss rates for massive particles. 
Excitation and ionization losses (i.e., Bethe-Bloch interactions; red bands) dominate for most energies of interest.
However, Bremsstrahlung (green band) becomes important for \bt-particles at very high energies, while plasma losses (blue bands) are dominant for \al-particles and fission fragments at low energies.
}
\label{fig:dedt_alli}
\end{figure}

The rate at which radioactive decay products transfer their energy to the ejecta depends on the ejecta composition.
Since our models produce different abundance patterns, these rates vary from slightly model to model, and we calculate for each model a set of energy-loss rates consistent with its composition.
The variation is generally small, however, as can be seen in Figure~\ref{fig:dedt_alli}, which shows the range of energy-loss rates predicted by our models.
The variance is usually within a factor of 2, and is often much less.
We summarize the relevant interactions below, but refer the reader to \citetalias{Barnes_etal_2016} (their \S3) for a detailed description.

\vspace{ 1mm }
\noindent\textbf{$\boldsymbol{\gamma}$-rays:}
The \gm-rays from radioactive decay lose their energy in photoionization and Compton scattering events.
The associated opacities are plotted in the top panel of Fig.~\ref{fig:dedt_alli}.
We constructed photoionization cross sections for each model's composition from the element-specific, energy-dependent cross sections available through the Photon Cross Section Database \citep[\emph{XCOM};][]{NIST_XCOM}, published by the National Institute of Standards and Technology (NIST).
Compton scattering cross sections were calculated analytically using the Klein-Nishina formula.
Photoionization opacity is much higher than Compton opacity at $E_{\gm} \lesssim 0.5$ MeV.
Both processes produce an electron, which often has an energy greater than the thermal energy of the background gas.
These secondary electrons are propagated through the ejecta until they have been effectively thermalized.

\vspace{ 1mm }
\noindent\textbf{$\boldsymbol{\beta}$-particles:}
Energetic \bt-particles transfer their energy by exciting or ionizing bound electrons, or through Coulomb interactions with free electrons.
The highest energy \bt-particles may produce Bremsstrahlung radiation that can itself thermalize or escape from the ejecta.
The importance of each of these energy-loss channels is presented in the second panel of Fig.~\ref{fig:dedt_alli}.

We model ionization and excitation losses---which dominate \bt-particle thermalization---with the well-established Bethe-Bloch formula \citep{Heitler_1954_qntmRad, Berger_Seltzer_1964, Gould_1975, Blumenthal_Gould_1970}, including relativistic corrections.
We use mass-weighted averages for composition-dependent quantities, such as the bound-electron number density and the average excitation energy.
Losses due to Coulomb interactions with unbound thermal electrons are described by the formulae of \citet{Huba_NRL}.
We adopt the formula appropriate in the limit that the energy of the thermalizing particle greatly exceeds the energy of thermal electrons (as in Eq. 4 of \citetalias{Barnes_etal_2016}).
Finally, the Bremsstrahlung stopping is computed following  \citet{Seltzer_Berger_Brem_1986}, which requires the use of $Z$-dependent empirical fitting constants. 
We use constants corresponding to the mass-weighted average atomic number of each model's composition.

\vspace{ 1mm }
\noindent\textbf{$\boldsymbol{\alpha}$-particles:}
Like \bt-particles, \al-particles thermalize via interactions with background thermal electrons.
In the case of free electrons, these interactions can be modeled using the formula of \citet{Huba_NRL} describing the energy lost by a suprathermal ion to thermal electrons in a plasma. 
The energy lost to bound electrons is calculated from \al-particle stopping powers retrieved from NIST's Stopping Power and Range Tables for Helium Ions \citep[\emph{ASTAR};][]{NIST_*Star}.
As in \citetalias{Barnes_etal_2016}, we map the full composition of each model to a simplified composition comprised of elements for which \al-stopping data is available.
Alpha-particle energy-loss rates are plotted in the third panel of Fig.~\ref{fig:dedt_alli}

\vspace{ 1mm }
\noindent\textbf{Fission fragments:}
In addition to free and bound electrons, fission fragments thermalize through Coulomb interactions with atomic nuclei.
Fission fragment thermalizaton is complicated by the question of ion state; fragments are not fully ionized, and their charge state $Z_{\rm ff, eff}$ depends on their kinetic energy. 
We determine $Z_{\rm ff, eff}$ using the formula of \citet{Schiwietz_Grande_2001_chargeState}.

We can then calculate losses due to free electrons using the same ion-electron formula as for \al-particles, with terms involving ion mass and charge updated appropriately. 
As in \citetalias{Barnes_etal_2016}, in the case of free electrons only, we set a floor $Z_{\rm ff, eff} \geq 7$, motivated by the fact that low ionic charge states permit thermal electrons to impact the fragment at distances smaller than the fission fragment radius.
To model interactions of fission fragments with bound electrons, we follow the procedure of \citet{Ziegler_1980} and scale proton stopping powers, which we take from NIST's \emph{PSTAR} database \citep{NIST_*Star}, by $Z^2_{\rm ff, eff}$.
The energy loss from interactions between fission fragments and atomic \emph{nuclei} is given by the nuclear stopping formula of \citet{Ziegler_1980}; it is subdominant for the energies of interest, as is apparent in the bottom panel of Fig.~\ref{fig:dedt_alli}.

\subsection{Particle Transport}

Radioactive particles are emitted in the ejecta at a rate proportional to the local mass density.
We assume that the ejecta has a uniform composition, so radioactive decay does not depend on position within the ejecta.
(In reality, thermodynamic conditions are likely to vary within an outflow, resulting in spatially dependent patterns of nucleosynthesis and radioactive decay.
We defer an exploration of these effects to later work.)
As decay products traverse the ejecta, either following magnetic field lines (in the case of charged fission fragments, \mbox{\al-,} and \bt-particles) or undergoing discrete scattering and absorption events (in the case of \gm-rays), the energy they transfer to the ejecta is calculated and tallied up.

Unlike in \citetalias{Barnes_etal_2016}, we track the deposition of radioactive energy in both time and space, enabling a more detailed description of thermalization. 
It has long been recognized that, since thermalization depends on density, its efficiency should vary within the ejecta \cite[e.g.][]{Wollaeger.ea_2018MNRAS_knova.lc.morph}.
In particular, since the densest regions will emit more radioactive energy, thermalize that energy more efficiently, and remain optically thick out to later epochs than less dense regions,
spatially dependent thermalization can impact the the temperature of the ejecta's photosphere, and thus the overall brightness and effective temperature of the kilonova (see \citet{korobkin.ea_2020.arxiv_axisym.kn.radtran} for a nice discussion).

\subsection{Ejecta Model}\label{subsec:modej}
To better highlight the variation in light curve predictions arising from the choice of nuclear physics parameters, we consider in this study a single ejecta model.
Our model is spherically symmetric with a broken power-law density profile, $\rho(v) \propto v^{-\eta}$, with $\eta = 1$ ($10$) in the inner (outer) ejecta layers, and the transition point set by the requirement that the mass distribution yields the specified ejecta mass and kinetic energy.
Motivated by the inferred mass and velocity of the red component of the GW170817 kilonova \citep{Kasen.ea_2017Natur_gw.170817.knova.theory,Drout.ea_2017Sci_gw.170817.emcp.disc,Villar.ea_2017ApJ_gw.170817.knova.agg,Tanaka.ea_2017PASJ_gw.170817.knova.interp,Perego.ea_2017ApJ_gw.170817.knova.model.3comp}, we set $M_{\rm ej} = 0.04 \msun$, and $v_{\rm ej} = 2(E_{\rm kin}/M_{\rm ej})^{1/2} = 0.1c$, with $c$ the speed of light.

Magnetic field configuration was shown in \citetalias{Barnes_etal_2016} to affect thermalization efficiency.
Here, we assume the magnetic field lines are ``tangled'' as a result of turbulent processes early in the ejecta's expansion history.
From a thermalization standpoint, this represents a middle ground between radial fields (which escort charged particles efficiently out of the ejecta, reducing the time during which the particles experience thermalizing interactions) and toroidal fields (which hold the charged particles in the ejecta's interior, maximizing thermalization).

\section{Thermalization Results}\label{sec:therm_results}

\begin{figure*}\includegraphics[width=\textwidth]{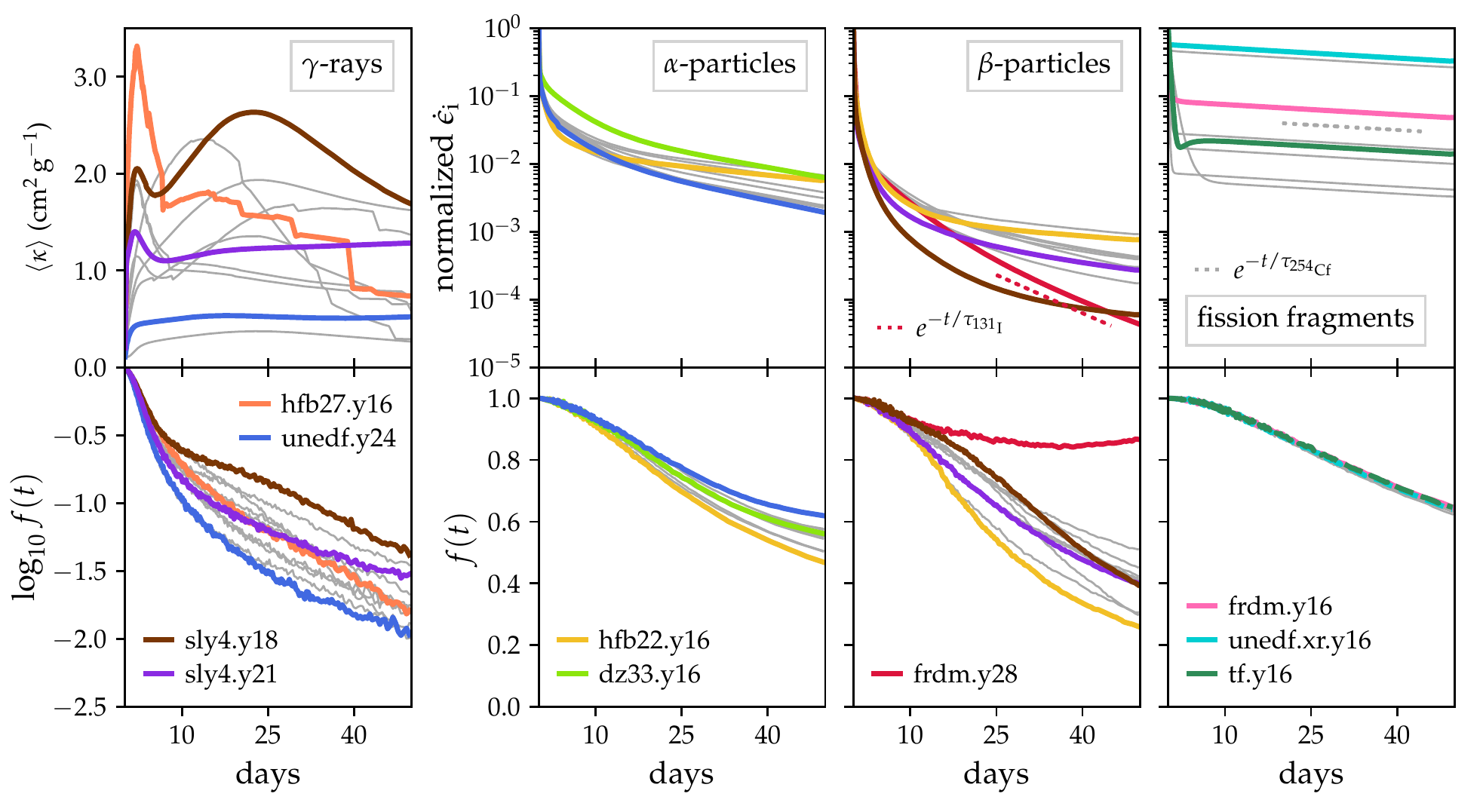}
\caption{Thermalization efficiencies of all decay products for all models (gray curves), with select cases color-coded to illustrate relevant trends.
The full data set is provided in Appendix \ref{appx:ft_all}. 
\textit{Top panels, left to right:} The spectrum-weighted \gm-ray opacity, and the rates at which radioactive energy is released as \bt-particles, \al-particles, and fission fragments. 
To facilitate comparison among models, we normalize all heating rates such that $\edoti = 1$ at $t = 0.1$ days. 
In certain cases (\bt-heating for \mtwo{} and fission heating generally) the absolute heating is dominated by a single isotope. 
Arbitrarily normalized exponential heating curves for these isotopes are plotted as dotted lines.
\textit{Bottom panels:} Thermalization efficiencies for each decay product. 
Gamma-ray thermalization is determined primarily by the effective \gm-ray opacity of the ejecta, with higher opacities raising $f_\gm(t)$.  
Massive particle thermalization is most sensitive to the slope of \edoti; the more steeply \edoti{} declines, the higher $f_{\rm i}(t)$.}
\label{fig:ft_all}
\end{figure*}

We define the time-dependent thermalization efficiency $f(t)$ as the ratio of the rate at which energy is absorbed by the ejecta (i.e., \emph{thermalized}) to the rate at which it is produced by radioactive processes.
To clarify the role of different decay channels and decay products on the overall thermalization, we calculate $f_{\rm i}(t)$ separately for each decay product $i$. 
For example, $f_{\al}(t)$ is the rate at which \al-particle kinetic energy is transferred to the thermal pool, divided by the rate at which radioactivity injects energy into the ejecta in the form of suprathermal \al-particles. 
The thermalization efficiencies for all decay products (barring neutrinos, which do not thermalize) are presented in the bottom panels of Figure~\ref{fig:ft_all}.
We have highlighted select models that illustrate important trends in thermalization, and plotted the remaining models as gray lines. 
However, $f_{\rm i}(t)$ for the full subset of models are included in Appendix \ref{appx:ft_all}.

The pattern most apparent in Fig.~\ref{fig:ft_all} is the dependence of thermalization on particle type.
As expected from earlier work \citepalias{Barnes_etal_2016} and analytic considerations \citep{Kasen.Barnes_2019_analytic.therm.knova}, we find that \gm-rays thermalize only feebly, \bt- and \al-particles thermalize more strongly, and fission fragments thermalize most robustly.
However, except in the case of fission fragments (which we discuss in more detail in \S \ref{subsec:ft_fiss}), there is also significant variation in $f_{\rm i}(t)$ among the different nuclear physics models.
The range of results for $f_{\rm i}(t)$ demonstrate the advantage of a comprehensive numerical treatment of \rp\ radioactivity to understand heating in kilonovae.

\subsection{Effect of absolute heating rate}
\label{subsec:ei_ft}
Much of the variability in $f_{\al}(t)$ and $f_\bt(t)$ can be explained by differences in the radioactive energy generation rates, $\dot{\epsilon}_{\rm i}$, which are shown in the top middle panels of Fig.~\ref{fig:ft_all}, normalized to $\dot{\epsilon}_{\rm i} =1$ at $t = 0.1$ days to aid comparison.
We find that a more steeply declining \edoti\ corresponds to a shallower decrease in $f_{\rm i}(t)$, and therefore to higher levels of thermalization.
(Of course, since less energy is emitted in the steeply declining cases, the total \emph{amount} of thermalized energy is still generally lower.)

The effect of \edoti\ is strongest at later times, after thermalization has started to become inefficient.
Prior to that, particles thermalize easily and $f_{\rm i}(t)$ hovers near unity regardless of the rate of decline.
In realistic models of the \rp, where the energy generation rate changes with time, the slope at later times will have a stronger effect on thermalization than the early-time decline.

This trend is illustrated by the \bt-particle thermalization of \mnine{} (model 9). 
At early times, the \bt-heating of this model, $\dot{\epsilon}_{\beta,9}$, falls rapidly, and $f_{\bt,9}(t)$ is high compared to the full set of models.
But by $t \sim 10$ days, $\dot{\epsilon}_{\beta,9}$ has flattened, becoming less steep than the \bt-heating rate of \mtwo{} (model 2).
Around the same time, $f_{\beta,9}(t)$ diverges from $f_{\beta,2}(t)$, falling to lower values while the latter retains its shallow slope.

Such behavior is expected from analytical models \citep[e.g.,][Eq. 26]{Kasen.Barnes_2019_analytic.therm.knova}, and can be understood by considering the energy evolution of a population of particles generated by radioactive decays.
The energy thermalized at any time $t$ comes from a collection of suprathermal decay products emitted at times ${\leq}t$.
The oldest unthermalized particles will, on average, have the lowest energies
and the the lowest thermalization times, since to zeroth order thermalization time scales with particle energy. When \edoti\ declines steeply, the fraction of the total unthermalized energy carried by older, rapidly-thermalizing particles increases, which flattens the slope of $f_{\rm i}(t)$.

The decay rate for single-isotope heating, $\edoti \propto e^{-t/\tau_{\rm r}}$, with $\tau_{\rm r}$ the isotope lifetime, will generically be steeper than the rate from an ensemble of nuclei \citep{Li_Paz_1998}.
Analytic thermalization models suggest that, for exponential decay, $f_{\rm i}(t)$ will pass through a local minimum and begin to increase with time.
While the sheer number of nuclei produced make truly exponential decay unlikely for the \rp, in some instances the heating may be dominated by one or a handful of nuclei, and become approximately exponential in nature.

Nucleosynthesis in high-\ye\ conditions burns a narrower distribution of nuclei, and so is more likely to become dominated by a single decay chain and transition into the regime of quasi-exponential decay.
Indeed, the three models with the highest initial \ye{} (models \mtwo, \meig, and \mten, with $\ye = 0.28,\, 0.24$, and $0.21$, respectively) derive their energy from a much smaller set of nuclear decays than other models.
However, the decrease and late-time increase in $f_{\rm i}(t)$ characteristic of exponential heating is found only in the \bt-particle thermalization of \mtwo\ (model 2).
As shown in Fig.~\ref{fig:ft_all}, $f_{\bt,2}(t)$ reaches a minimum of 0.84 at $t = 36$ days, then rises to 0.87 by $t = 50$ days, a value ${\gtrsim}70\%$ greater than the next-highest model.
This behavior is the result of the nearly exponential form of $\dot{\epsilon}_{\bt,2}$.

\begin{figure}\includegraphics[width=\columnwidth]{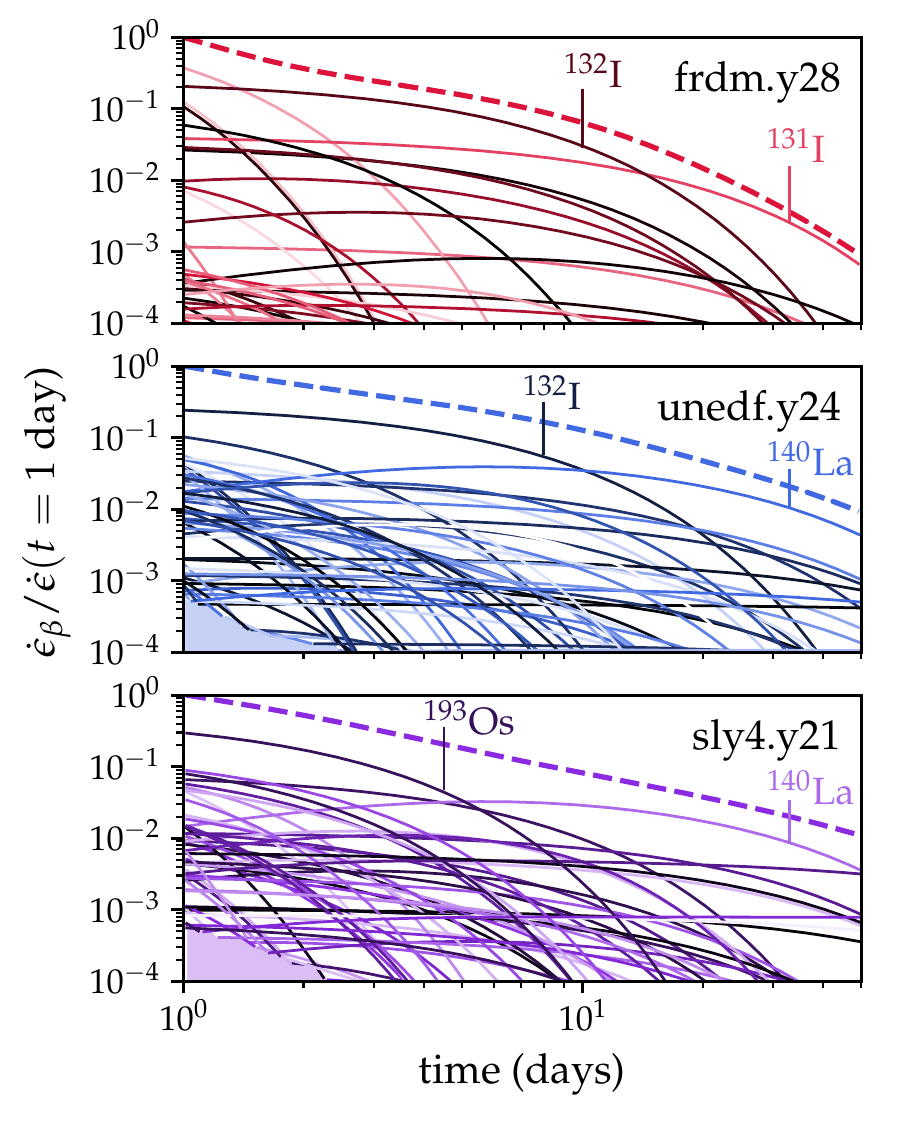}
\caption{The energy released in \bt-particles for the three models with high initial \ye\ (\mtwo, \meig, and \mten) for $t \geq 1$ day. 
The thick, dashed lines show the total heating rate, while the heating from individual isotopes are plotted in color scale. 
We have normalized all rates such that $\dot{\epsilon}_{\bt, \rm tot} = 1$ for each model at $t = 1$ day.
Our spectrum calculation includes only the top 50 \bt-decaying nuclei, so the early-time contributions of minor nuclei have sometimes not been recorded. 
Transparent solid patches in the middle and bottom panels mark the region of parameter space for which decay data is missing.
The total \bt-particle heating for \mtwo\ most closely approximates an exponential due to the dominance of \iso{I}{132} (at $2 \leq t/{\rm day} \leq 15$) and especially \iso{I}{131} ($t/{\rm day} \geq 15$) on timescales close to these isotopes' lifetimes.}
\label{fig:beta_isos}
\end{figure}

This is shown in Figure~\ref{fig:beta_isos}, which plots the total absolute \bt-particle heating (dashed lines) for the three high-\ye{} models, as well as the heating from individual isotopes (solid lines), all normalized to the total \bt-heating rate at $t=1.0$ day.
By $t = 15$ days, the heating for \mtwo{} (model 2) tracks the decay of \iso{I}{131}, which is responsible for ${\sim}70-80\%$ of $\dot{\epsilon}_{\bt,2}$ thereafter.
The decay timescale of \iso{I}{131} is 8.03 days, and its production, in our simulations, occurs at $t \leq 1$ day.
Thus, by the time \iso{I}{131} dominates the total heating, it is well into its decay phase, and can impart to $\dot{\epsilon}_{\bt,2}$ its steep exponential decline (see also the top third panel of Fig.~\ref{fig:ft_all}).

The relationship between \edoti\ and $f_{\rm i}(t)$ slightly attenuates the variability in effective heating.
Models with more steeply declining \edoti\ will produce less energy through radioactivity, but will thermalize that energy more effectively.
However, as we will show in \S\ref{subsec:ft_net}, the effect is not strong enough to overcome the underlying differences in absolute heating.

\subsection{Effect of spectra}\label{subsec:speceffect}
While the form of \edoti\ explains some of the diversity in $f_{\rm i}(t)$, the magnitude and shape of the thermalization curves also depend on the energy spectra of the emitted particles, which varies among models and changes with time.
The importance of the spectrum is particularly apparent for \bt-particle and \gm-ray thermalization.

\begin{figure}\includegraphics[width=\columnwidth]{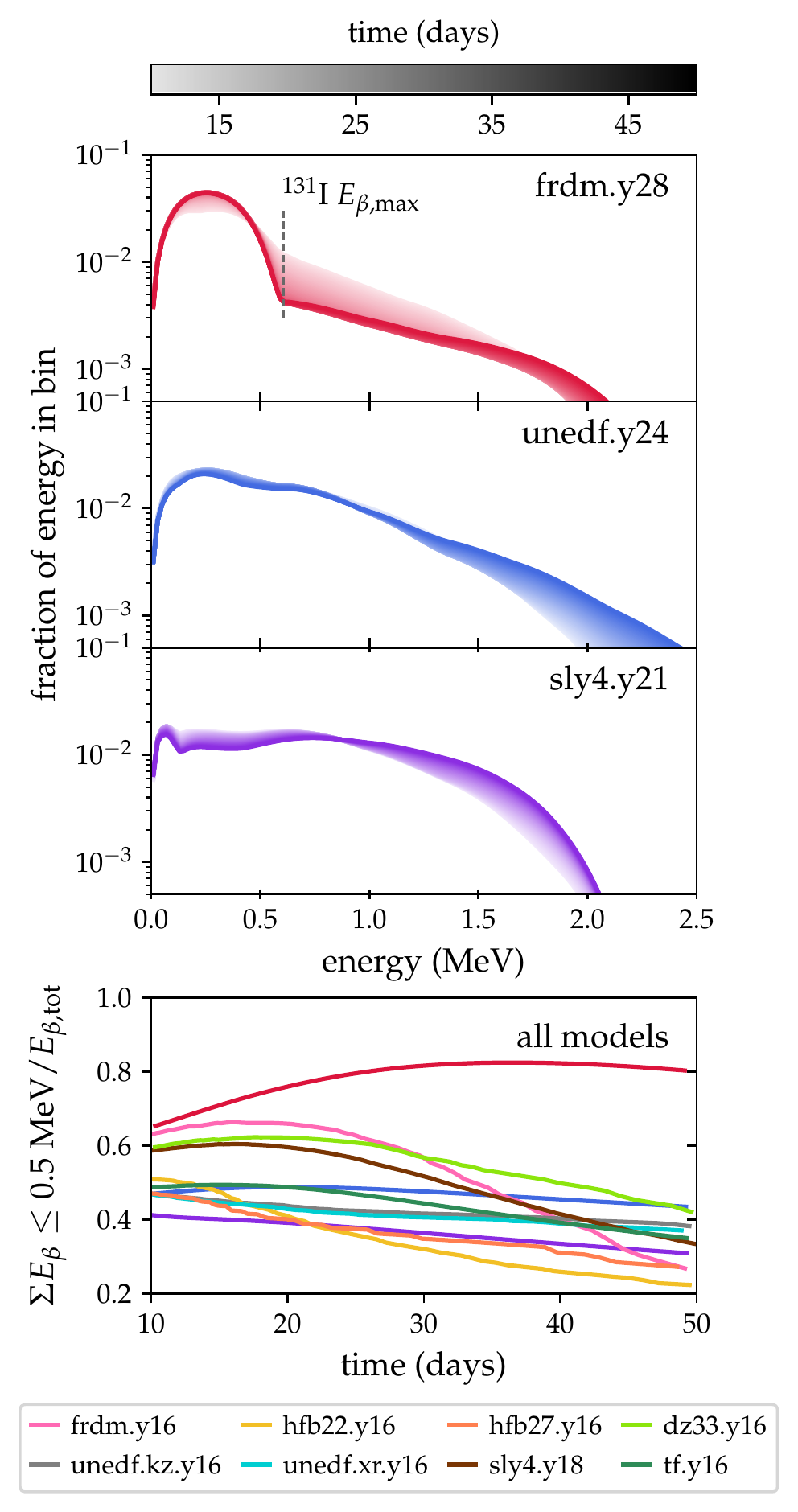}
\caption{ Differences in the \bt-emission spectra among models. 
\textit{Top three panels:} The later-time \bt-emission spectra of the high-\ye\ models \mtwo, \meig, and \mten.
Because its \bt-decay radioactivity is dominated by \iso{I}{131} at $t \gtrsim 15$ days, \mtwo{} has a very low-energy spectrum, which increases $f_\bt(t)$.
\textit{Bottom panel:} The fraction of emitted \bt-decay energy accruing to particles with $E_\bt \leq 0.5$ MeV for all models as a function of time. 
Model \mtwo{} (red curve) is exceptional in this regard, which in part explains its surprisingly high thermalization efficiency.
}
\label{fig:bspec_hiye}
\end{figure}

\vspace{ 1mm }
\noindent\textbf{$\boldsymbol{\beta}$-particles:} 
The role of the emission spectrum in \bt-particle thermalization is most obvious for \mtwo{} (model 2).
While the quasi-exponential shape of $\dot{\epsilon}_{\bt,2}$ explains the unique falling and rising behavior of $f_{\bt,2}(t)$, the sustained efficient thermalization is additionally due to an unusually low-energy \bt-spectrum. 
Figure~\ref{fig:bspec_hiye} compares the later time \bt-emission spectra of the high-\ye\ models (\mtwo, \meig, and \mten) and presents for all models the fraction of the total \bt-particle heating supplied by particles with $E_{\bt,0} \leq 0.5$ MeV, as a function of time.
The concentration of \mtwo's \bt-spectrum at lower energies makes it an outlier, even relative to the other high-\ye\ models, which generally produce lighter nuclei with lower decay energies.
This again is due to the predominance of \iso{I}{131}.
Almost 90\% of \iso{I}{131} decays proceed through the emission of \bt-particles whose maximum energy is 606 keV, putting the average energy of a \bt-particle from this decay at $\langle E_\bt \rangle = 191$ keV.
Such low-energy particles thermalize more quickly, enhancing $f_\bt(t)$.

Compared to other decay products, \bt-particles exhibit a high degree of variability in their thermalization.
Our findings indicate that even in the regime of high-\ye\ \rp\ nucleosynthesis, radioactivity and thermalization can vary substantially, and $f_\bt(t)$ is sensitive to the particular constellation of isotopes synthesized.

\vspace{ 1mm }
\noindent\textbf{$\boldsymbol{\gamma}$-rays:}
Gamma-ray thermalization occurs via a series of distinct scattering and absorption events, rather than the continual slowing-down processes that mediate the thermalization of massive particles. 
The probability of scattering or being absorbed is a function of the optical depth, $\tau_\gm = \int \rho \kappa_\gm \, \rm{d}s$, with $\rho$ the mass density, $\kappa_\gm$ the \gm-ray opacity, and $s$ the path length.

At energies $E_\gm \gtrsim 0.5$ MeV, $\kappa_\gm$ is set by the Compton scattering opacity, which has a fairly low value ($\kappa_{\rm Comp} \lesssim 0.1$ cm$^2$ g$^{-1}$) and is relatively flat with increasing $E_\gm$.
For $E_\gm \lesssim 0.5$ MeV, photoionization becomes important, providing an opacity $\kappa_{\rm PI}$ orders of magnitude greater than $\kappa_{\rm Comp}$, and rising sharply as $E_\gm$ decreases (Fig.~\ref{fig:dedt_alli}).
As a result, \gm-ray thermalization depends strongly on the emission spectrum of \gm-rays; if most \gm rays are emitted at energies for which $\kappa_\gm$ is high, $f_\gm(t)$ will be augmented.

To illustrate this, we calculate as a function of time a spectrum-weighted average \gm-ray opacity, 
\begin{align*}
    \kav = \sum \kappa_\gm(E) \, \phi(E),
\end{align*}
where $\phi(E)$ is the fraction of \gm-ray energy emitted in the range $E \leq E_\gm \leq E + \delta E$.
We show \kav\ for all models in the top-left panel of Fig.~\ref{fig:ft_all}.

While some of the model-to-model differences are due to composition (e.g., different abundance patterns resulting in slightly different net opacities) most of the variation among models, and all of the variation with time for any particular model, is due to differences in the \gm-emission spectrum.
Because $\kappa_\gm$ increases so steeply as $E_\gm$ falls below ${\sim}0.5$ MeV, \kav\ is much more sensitive to the spectrum at low $E_\gm$; low-energy \gm-rays can dominate \kav, and therefore thermalization, even when they represent only a small fraction of the total energy from \gm-ray emission.

As anticipated, \kav\ is strongly correlated with \gm-ray thermalization (lower left panel of Fig.~\ref{fig:ft_all}).
To the extent that \kav\ is not perfectly predictive of $f_\gm(t)$, the discrepancy is most likely due to a confluence of second-order effects, such as the rate-of-decline of $\dot{\epsilon}_\gm$, or the strength of the \gm-emission spectrum at energies just above the point where $\kappa_{\rm PI}$ begins to dominate the opacity.
Photons emitted at these energies could down-scatter into the photoionization regime, and thereafter thermalize more efficiently than their initial energies would suggest.
Since the ejecta is largely transparent to \gm-rays, as indicated by the overall low efficiencies, even a small number of such down-scattering events could represent a relatively large fraction of all thermalizing interactions, and therefore impact $f_\gm(t)$.

\subsection{Fission Fragments and Californium}\label{subsec:ft_fiss}

Unlike \gm-rays, \bt-particles, and \al-particles, fission fragment thermalization is uniform across models, seeming to counter the conclusion that $f_{\rm i}(t)$ is sensitive to the details of radioactivity. 
However, fission fragments are in fact the exception that proves the rule:
the results are consistent because our models' predictions of fission radioactivity---despite the range of fission treatments considered---are quite uniform, and because we assume a single fission fragment mass and flat energy distribution for all models.

At early times, thermalization is generally efficient, particularly for rapidly thermalizing fission fragments. 
But at times late enough for $f_{\rm ff}(t)$ to be sensitive to the slope of $\dot{\epsilon}_{\rm ff}$, only two isotopes contribute to fission heating.
The first, \iso{Cf}{254}, has a well-established half-life of 60.5 days \citep{Phillips.ea.Gatti_JInorgChem.1963_Cf254.halflife}.
It is present in all models save \mtwo{}, which has no measurable fission.
Its production by the \rp{} and its potential impact on kilonova light curves via thermalization were explored by \citet{Zhu.ea.Mumpower_2018_Cf254.knova}.
The second nucleus, \iso{Rf}{269}, appears for \mthree{} and \mfour{}.
There have been no experimental measurements of \iso{Rf}{269}, and its predicted fission half-life is sensitive to the theoretical models of nuclear masses and spontaneous fission lifetimes adopted.
The HFB models for which \iso{Rf}{269} is found to contribute substantively to fission heating have fission barriers computed for HFB14 and half-lives calculated according to \citet{Karpov.ea_2012ijmpe_heavy.elem.fiss} and \citet{Zagrebaev.ea_2011PRC_nrich.heavy.fiss}.
With these inputs, the spontaneous fission half-life of \iso{Rf}{269} is predicted to be very close to that of \iso{Cf}{254}: 58.95 days.
(The longevity of these isotopes, along with the overall efficient thermalization of fission fragments, keep $f_{\rm ff}(t)$ from reaching a local minimum within the time limit of our simulation, despite $\dot{\epsilon}_{\rm ff}$ being effectively exponential.)

\begin{figure}[htp]\includegraphics[width=\columnwidth]{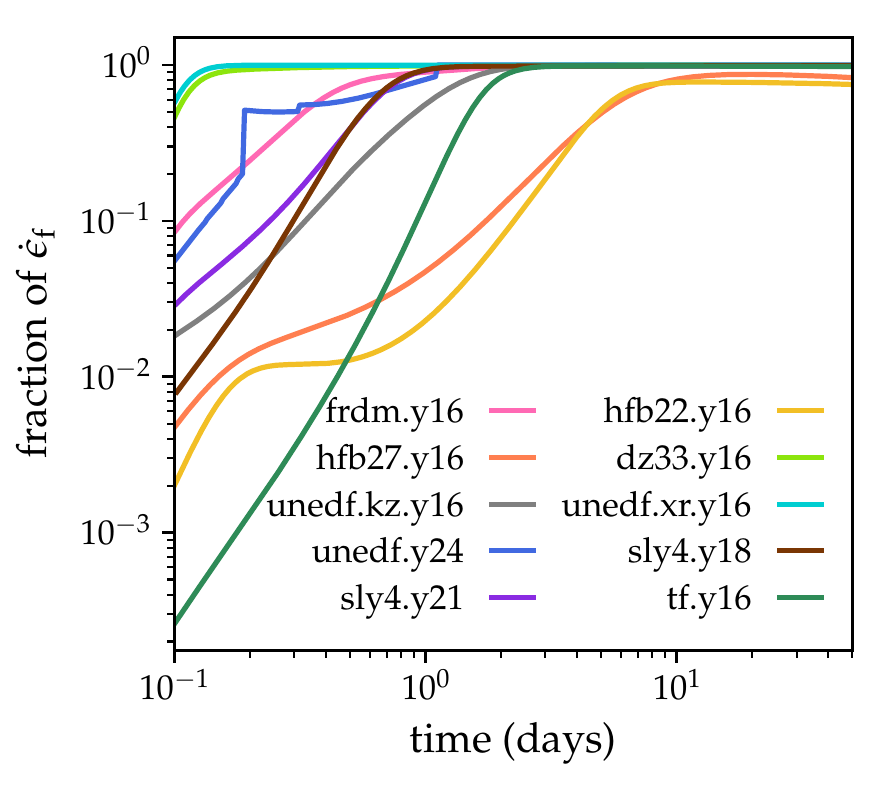}
    \caption{The fraction of total fission heating supplied by \iso{Cf}{254} and \iso{Rf}{269} for all models as a function of time. 
    After $t \approx 10$ days, nearly all fission is from either both these isotopes (\mthree{} and \mfour{}), or \iso{Cf}{254} alone (all other models).
    Their similar half-lives (60.5 days for \iso{Cf}{254} and 
    58.95 days for \iso{Rf}{269}) cause the fission heating curves for all models to have very similar slopes. Model \mtwo{} produced no fissioning isotopes, and so is omitted from this figure.}
    \label{fig:fiss_iso}
\end{figure}

\begin{figure*}[t]\includegraphics[width=\textwidth]{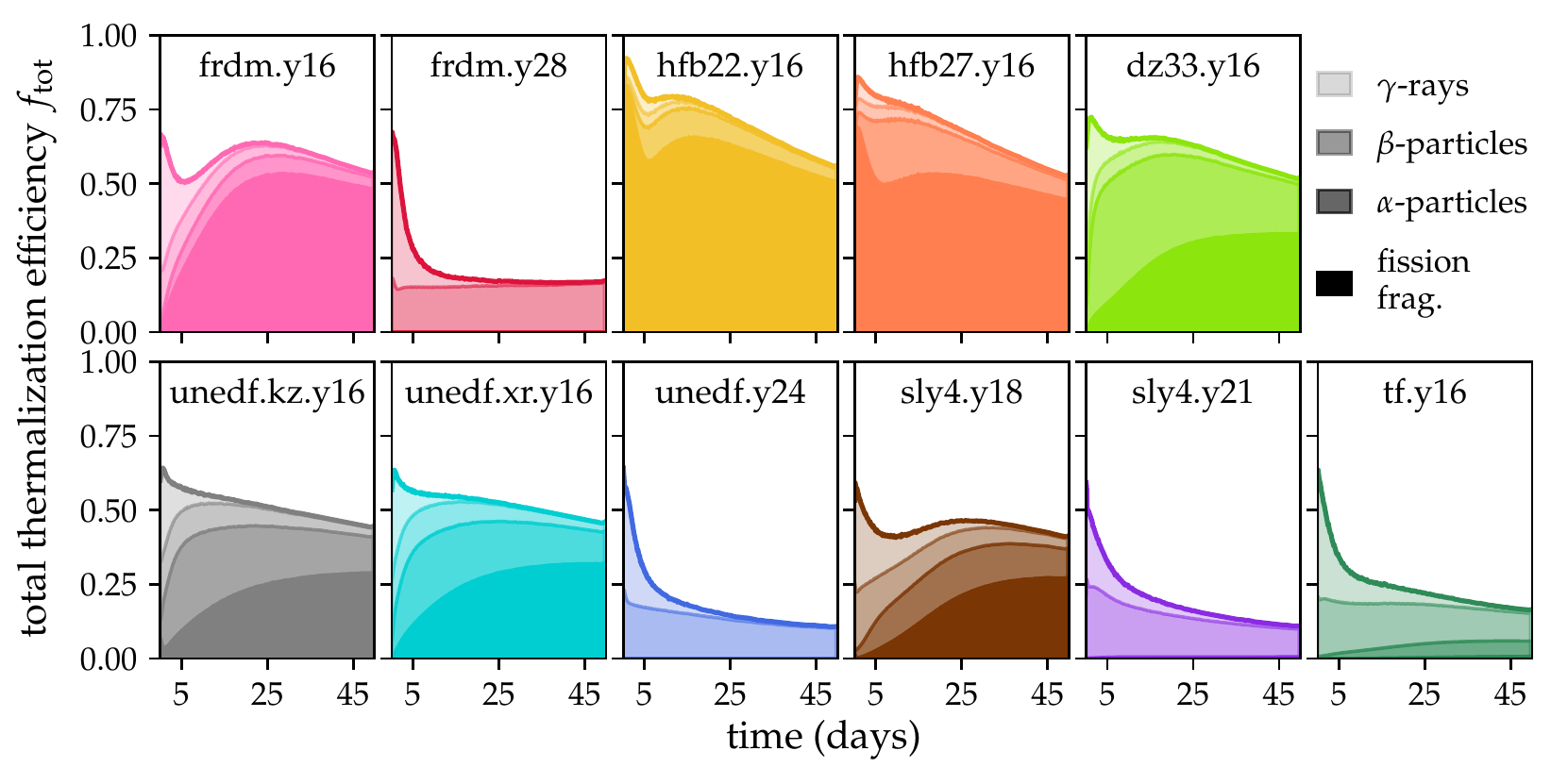}
\caption{The total time-dependent thermalization efficiencies $f_{\rm tot}(t)$ for all models, showing the contribution of the four decay products to the the total thermalized energy.
In models where \al-decay and fission are non-negligible, they dominate the thermalized energy at late times, causing $f_{\rm tot}(t)$ to flatten, or even to rise from a local minimum caused by the increasing inefficiency of \bt-particle thermalization.
Models without significant \al-decay and fission (\mtwo{}, \meig{}, and \mten{}) have overall lower levels of thermalization.
}
\label{fig:ft_tot}
\end{figure*}

Since nearly all fission after early times is due to these two isotopes with similar half-lives (Fig.~\ref{fig:fiss_iso}) and since we lack a detailed description of the fission fragment distribution, and so use the same fission spectrum for each case (\S \ref{subsec:radspec}), late-time fission in every model is, modulo normalization, effectively identical.
As a result, our results for $f_{\rm ff}(t)$ are similarly consistent.

However, this consistency depends on the theoretical calculation of the fission half-life and branching ratios of allowed decay modes of \iso{Rf}{269}. 
For example, 
some models populate \iso{Rf}{269}, but find \al-decay, rather than fission, to be its dominant decay mode.
A thorough experimental investigation of potentially long-lived fissioning isotopes could change this picture and allow
more accurate models of fission fragment thermalization and its effect on kilonova light curves.

\subsection{Net thermalization efficiency}
\label{subsec:ft_net}

The total thermalization efficiency from \rp\ decay depends on the efficiencies associated with individual particle types, as well as on the partition of radioactive energy into various decay channels.
In this section we present the net thermalization efficiency of each model, as well as the corresponding effective heating rates for the associated kilonova model.
Since thermalization depends on ejecta parameters, our results are valid only for the ejecta model described in \S \ref{subsec:modej}.

Figure~\ref{fig:ft_tot} shows the total thermalization efficiency for all models, highlighting the contribution of each particle type to the overall thermalization. 
Across all models, thermalization efficiencies range from 0.53 to 0.9 at early times, and decrease (though not necessarily monotonically) to between 0.1 and 0.5 at late times. 

The most conspicuous difference is between models that feature significant heating by \al-decay and fission---which have higher overall levels of thermalization---and models that not.
Even \mtwo's extremely high \bt-particle thermalization efficiency cannot compensate for its lack of rapidly thermalizing \al-particles and fission fragments. 
Since the radioactivity in \mtwo{} (and \meig{} and \mten{}) is dominated by \bt-decay, most of the energy is lost to neutrinos and \gm-rays, and even when \bt-particles themselves thermalize efficiently, this sets a restrictive upper limit on total thermalization.
In many cases where fission and \al-decay grow in importance, the net thermalization efficiency flattens dramatically, or even begins to rise, around $t = $ 10 -- 20 days, as a result of the more efficient thermalization associated with these decay modes.

\begin{figure*}\includegraphics{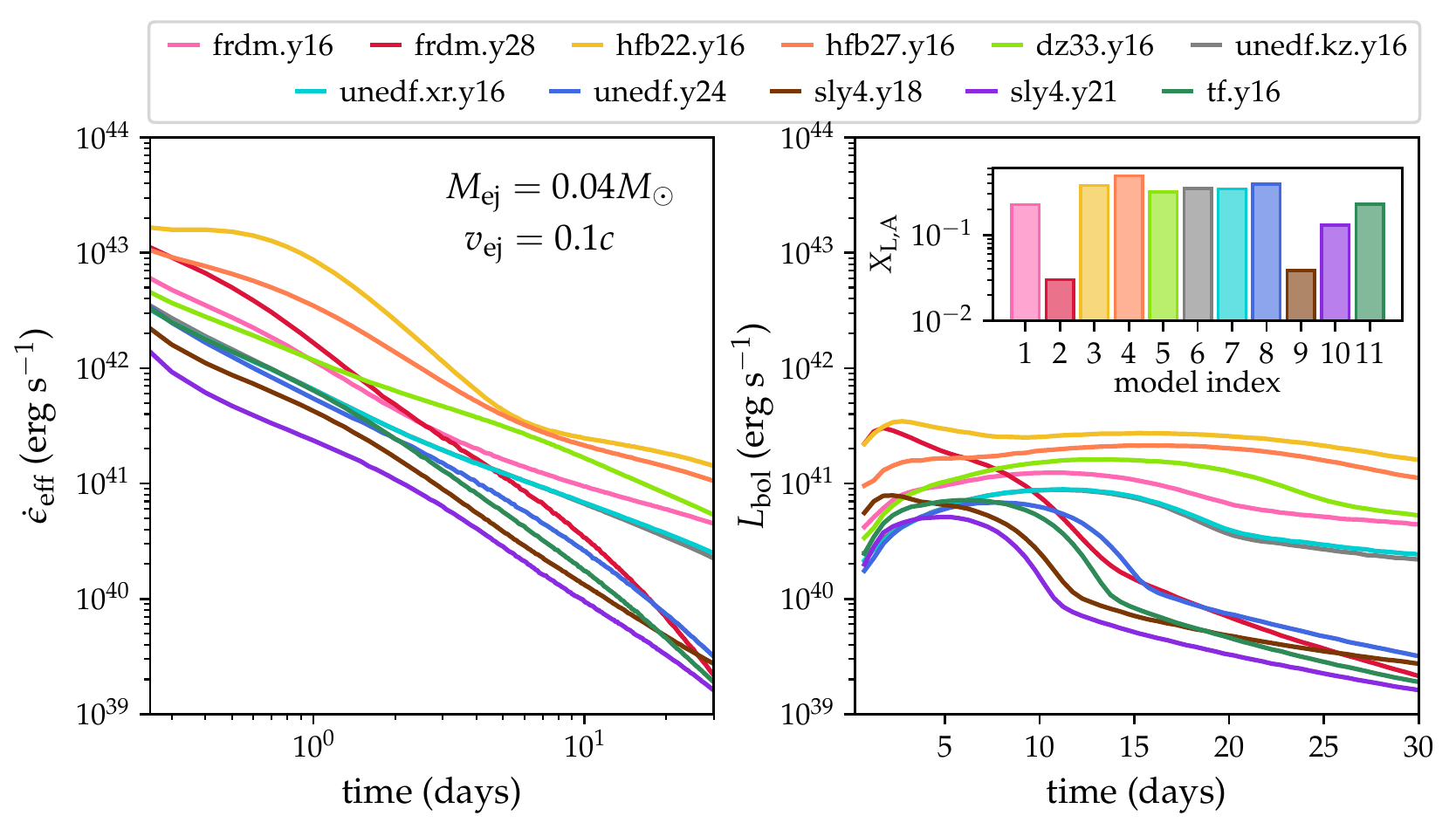}
    \caption{The effective heating rates and bolometric light curves of every model, for our chosen ejecta configuration $(\mej, \vej) = (0.04\msun, 0.1c)$.
    The $x$-axis of the left-hand (right-hand) panel is scaled logarithmically (linearly).
    \emph{Left panel: } Effective heating rates. There is considerable spread in both the magnitudes and the slopes of the curves, even at early times.
    We present the total heating rate (erg s$^{-1}$) rather than the specific heating rate (erg s$^{-1}$ g$^{-1}$) because thermalization depends on ejected mass and velocity, and therefore specific effective heating rates cannot be trivially scaled with \mej. \emph{Right panel: } Bolometric luminosity, with the combined lanthanide and actinide mass fraction for each model presented in the inset. The variability of $L_{\rm bol}$ matches that of $\dot{\epsilon}_{\rm eff}$. Lanthanide and actinide mass fraction is not entirely predictive of the light-curve width and rise time; rather, the shape of the light curve is determined by $X_{\rm L,A}$ in combination with $\dot{\epsilon}_{\rm eff}$.  }
    \label{fig:heff_lbol}
\end{figure*}

In the right-hand panel of Figure~\ref{fig:heff_lbol}, we plot the rate at which radioactive energy is deposited in the ejecta (i.e., $\eeff{}(t) = \mej \; \dot{\epsilon}_{\rm abs}(t) \times f(t)$) for all models.
These curves exhibit a degree of variation comparable to that of the absolute heating rates (Fig.~\ref{fig:model_qandy}), though the effects of model-specific thermalization have adjusted the shapes and magnitudes of the curves.
Critically, for $0.5$ days $\leq t \leq 15$ days, the time over which most kilonovae are expected to peak and fall, there is a difference of more than an order of magnitude separating the models with the lowest and highest heating.
This indicates that the kilonovae corresponding to the models in our subset will have a similarly wide range of luminosities.

\section{Kilonova Emission}\label{sec:kilonova}

We use the results of our thermalization calculations as inputs for radiation transport simulations, which we carry out with the time-dependent Monte Carlo radiation transport code \texttt{Sedona} \citep{Kasen_MC, Kasen.ea_SedonaCode_in.prep}.

\subsection{Details of the Radiation Transport Models}\label{subsec:sedona_mods}
We compute light curves for the same single ejecta model described in \S\ref{subsec:modej}.
We have modified \texttt{Sedona} so the heating rate can be specified in time and space by a discrete data set $\{\eeff{}(v, t)\}$, where $v$ is the ejecta velocity coordinate.
This allows us to import our effective heating rates directly into \texttt{Sedona}, instead of relying on best-fit functional forms to approximate \eeff{}.
It also enables us to capture the variation of \eeff{} in space. 
This is important because, as discussed in \citet{korobkin.ea_2020.arxiv_axisym.kn.radtran}, regions of higher mass density will emit more radioactive energy and better thermalize that energy.
This results in further stratification of the ejecta temperature: hot areas stay hotter and relatively cool areas cool faster.

The composition of the outflow is a critical input for kilonova modeling \citep{Even.ea_2020.apj_comp.effects.kne} because it sets the bound-bound opacity of the ejecta.
Because our atomic data set does not include data for all species in the periodic table, we derive from our \texttt{PRISM} results a modified composition for use in our \texttt{Sedona} calculations.
The major determinant of an element's opacity is its position in the periodic table: $f$-block elements have the highest opacity, followed by $d$-block, $p$-block, and $s$-block elements \citep{Kasen_2013_AS,Tanaka_Hotok_rpOps}.
However, opacity also varies within blocks.
Elements near the center of a periodic table row contribute a higher opacity due to their greater atomic complexity \citep{Tanaka.ea_2019_mnras_system.rproc.opacs}.

The simplified compositions are constructed to preserve, as closely as possible, the distribution of the highest-opacity species within their periodic table row.
We use the same atomic data set as in \citet{Kasen.ea_2017Natur_gw.170817.knova.theory}. It includes synthetic atomic data for all lanthanides except Lu ($Z=71$), but lacks actinide data.
In our model compositions, the mass fraction of any lanthanide is the mass fraction $X^{\rm P}$ of the element predicted by \texttt{PRISM}, plus the predicted mass fraction of the corresponding actinide.
For example, $X^{\rm mod}_{\rm Nd} = X^{\rm P}_{\rm Nd} + X^{\rm P}_{\rm U}$. 
Since we have no data for Lu, we send both Lu and its atomic structure homologue Lr ($Z=103$) to Yb ($Z=70$).
The mass fractions of the $d$-block elements are split evenly among $Z=21 - 28$, while $Z=20$ stands in for the $s$- and $p$-block remainder.
For these lighter elements, we use atomic data from the \texttt{CMFGEN} database \citep{Hillier.Lanz_2001.ASPC_cmfgen.code}, as in \citet{Kasen.ea_2017Natur_gw.170817.knova.theory}.

\subsection{Bolometric Light Curves}\label{subsec:lc_eeff}
Simple analytic considerations reveal how a light curve's time-to-peak, $t_{\rm pk}$, depends on key properties of the outflow,
\begin{equation}
    t_{\rm pk} \propto (\mej \kappa/\vej c)^{1/2},
    \label{eq:tpeak}
\end{equation}
where $\kappa$ is an effective (gray) opacity.
In \rp\ ejecta, opacity is dominated by lanthanide and actinide elements \citep{Kasen_2013_AS,Tanaka_Hotok_rpOps}, and $\kappa$ increases with \xlan{}.
The scalings of Eq.~\ref{eq:tpeak}, along with the rule of thumb $L_{\rm pk} \approx \eeff{}(t_{\rm pk})$ \citep[``Arnett's Law'';][]{Arnett_1980, Arnett_1982_Sne}, define a straightforward relationship between ejecta parameters, peak time, and peak luminosity.

Were \eeff{} consistent across our model suite, this relationship would account entirely for variability in our models' light curves: with \mej\ and \vej\ held constant, light-curve evolution would depend only on composition, and models with higher \xlan{} would have later-peaking, dimmer light curves \citep[e.g.][]{Barnes_2013}.

As can be seen in the right-hand panel of Fig.~\ref{fig:heff_lbol}, our models do exhibit a range of peak times and luminosities.
The diversity in predicted \eeff\ (left-hand panel) propagates through to the light curves; luminosities for the single ejecta model we consider differ by ${\sim}1$ order of magnitude at peak and by even more on the light-curve tail. 
However, because our models vary both \eeff\ and \xlan{} (re-plotted in the inset axes of Fig.~\ref{fig:heff_lbol}), they do not reproduce the expected correlation between $t_{\rm pk}$, $L_{\rm pk}$, and $\kappa$.

Most obviously, diversity in \eeff\ disrupts the relationship between $t_{\rm pk}$ and $L_{\rm pk}$.
The effect of Arnett's Law can still be seen in certain cases, although its influence is diminished relative to a scenario with uniform 
\eeff{}.
For example, the comparatively high luminosity of \mnine{} (model 9) at $t \lesssim 3$ days, is a product less of its effective heating than of its low opacity, which allows a larger fraction of \eeff{,9} to escape the ejecta on short timescales.
However, for the most part, the light curves with the most luminous peaks are those with the most energetic \eeff{}, rather than the shortest rise times.

A more subtle effect is that of the shape of \eeff\ on the kilonova rise time.
While low-\xlan{} cases (\mtwo, \mnine, and \mten) generally rise faster and sustain shorter photospheric phases than their high-\xlan{} counterparts (e.g., \msix{} and \msev),
not all models adhere to the expectation that peak time increases with \xlan{}.
Model 5 (\mfive), for instance, peaks later and transitions from peak to tail more slowly than \meig{} (model 8), despite having a lower \xlan{} ($\xlan{5} = 0.31 < \xlan{8} = 0.38$).

This is due to the shallower slope of \eeff{,5}.
When \eeff\ declines more slowly, the energy supplied by radioactivity more effectively offsets the loss, due to adiabatic expansion and diffusion of radiation, of energy deposited earlier, causing the light curve to evolve more slowly.
(Heat from a flatter \eeff\ also allows the ejecta to stay singly ionized longer, which enhances the opacity; see \S\ref{subsec:doublepk}.)
If \eeff\ remains shallow as the system becomes optically thin, it will also reduce the luminosity difference between the light-curve peak and tail.
(The largest anomaly in the \xlan{}--$t_{\rm pk}$ relationship appears to be \mthree, with $\xlan{} = 0.37$, and $t_{\rm pk} \approx 3$ days.
However, \mthree's quick rise results from a unique set of circumstances, as explained in \S\ref{subsec:doublepk}.)

\begin{figure*}\includegraphics{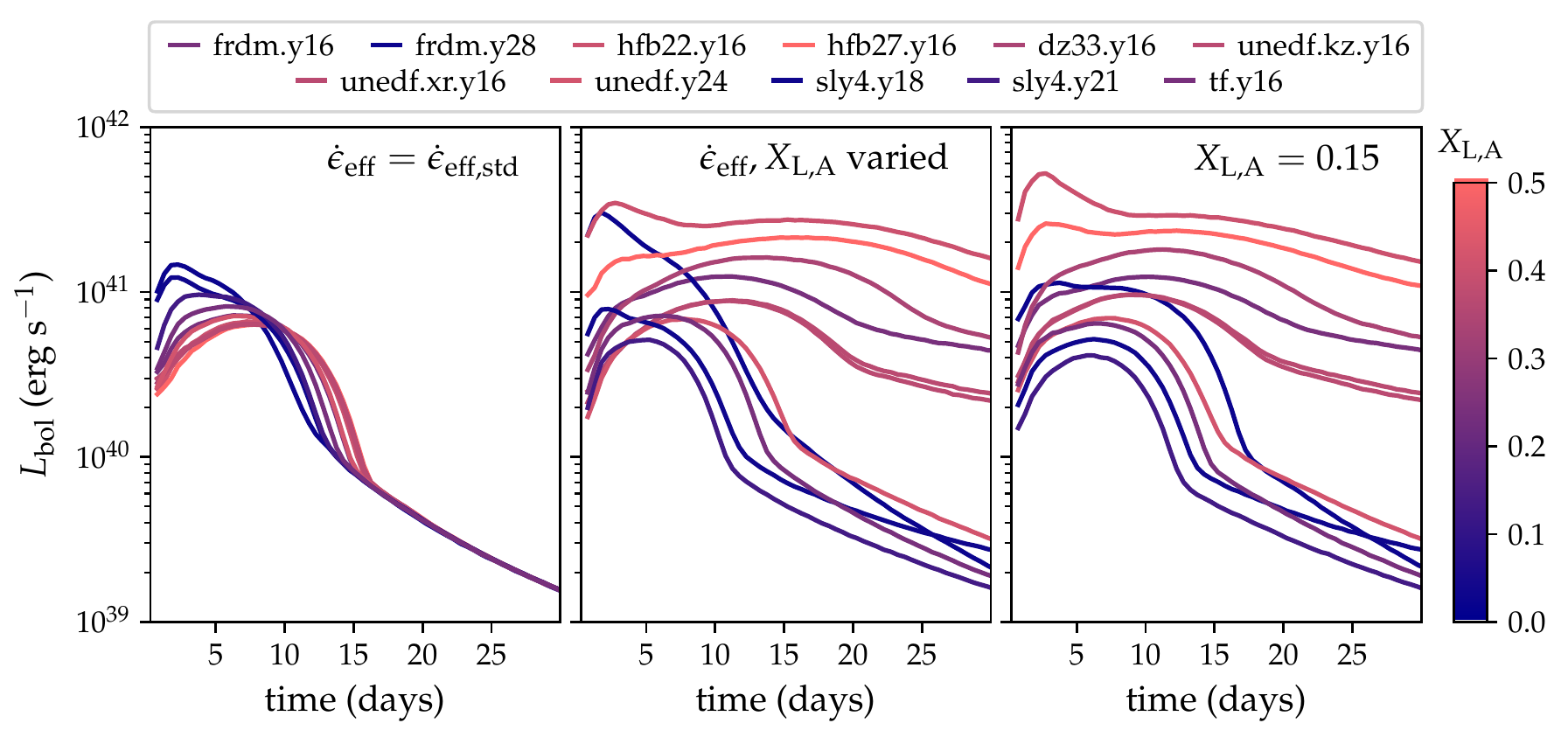}
    \caption{Three sets of light curve simulations that isolate the impacts of lanthanide and actinide mass fraction \xlan{} and the effective heating rate \eeff{}. 
    Unlike in other figures, the curves here are color-coded based on each model's (original) \xlan{}.
    \emph{Left panel:} Light curves that use a standard \rp\ heating rate, but preserve each model's composition. 
    The peak time ($t_{\rm pk}$) and peak luminosity ($L_{\rm pk}$) of each light curve are determined by \xlan{}.
    \emph{Middle panel:} The full set of models, with both \eeff\ and \xlan{} allowed to vary (i.e., the same light curves as in Fig.~\ref{fig:heff_lbol}).
    These curves do not reproduce the correlation between \xlan{}, $t_{\rm pk}$, and $L_{\rm pk}$ seen in the left-hand panel.
    \emph{Right panel:} Light curves for a modified suite of models in which the composition is uniform, but \eeff\ varies. (The color-coding is preserved to aid comparison; all light curves here are for models with $\xlan{} = 0.15$.)
    The shape of \eeff\ has a strong effect on the light-curve evolution independent of \xlan{}.
    }
    \label{fig:lcs_xe_std}
\end{figure*}

To more clearly disentangle the effects of \eeff\ and \xlan{} on light curves, we ran two sets of additional calculations.
In the first, we adopted a standard \rp\ heating rate, \eeff{,std}, similar to that of 
\citet{Kasen.ea_2017Natur_gw.170817.knova.theory}, but retain the compositions predicted by \texttt{PRISM}. 
In the second, we fixed the composition and set $\xlan{} = 0.15$, but use \eeff\ from our thermalization calculations.
The light curves of the uniform heating (composition) models are shown in the left-hand (right-hand) panel of Figure~\ref{fig:lcs_xe_std}, while the middle panel displays the fully consistent light curves of Fig.~\ref{fig:heff_lbol}.
For Fig.~\ref{fig:lcs_xe_std}, we employ an alternate color scheme in which the colors of the curves reflect each model's original \xlan{}.

As the left-hand panel of Fig.~\ref{fig:lcs_xe_std} demonstrates, when \eeff\ is fixed, \xlan{} is tightly correlated with $t_{\rm pk}$ and $L_{\rm pk}$.
However, adjusting \eeff{} (middle panel) confuses this pattern, and produces a more diverse set of light curves than would be expected from variation in \xlan{} alone.
A comparison of the middle and right-hand panels shows that the bulk of the diversity in our model light curves---in terms of both brightness and light-curve morphology---is due to differences in \eeff{}, rather than \xlan{}.
The light curves of the middle panel, which reflect the full diversity in nucleosynthesis and radioactivity predicted by our \texttt{PRISM} and thermalization simulations, clearly exhibit the greatest variety and most complex relationship between \xlan{}, \eeff{}, and the light-curve evolution.

Effective heating \eeff\ thus joins \mej, \vej, and $\kappa$ as a critical determinant of kilonova light curves, which introduces new complexities into the modeling and interpretation of kilonova emission.
For example, a plateau-shaped light curve could reflect a shallow-sloped \eeff\ as much as the high \mej, low \vej, and/or high opacity that Eq.~\ref{eq:tpeak} would suggest. 
Likewise, if \eeff\ is higher than is typically assumed, a given luminosity may be achieved by a lower \mej\ than would otherwise be required. 
Among our models, higher \eeff\ usually signals significant heating from fission,
which is correlated with both greater \xlan{} (since fission requires robust heavy-element nucleosynthesis) and flatter \eeff{}, due to the efficient thermalization of fission fragments and the long half-lives of the dominant fission reactions.
One or both of these mechanisms could extend the light curve, compensating for the decrease in optical depth and $t_{\rm pk}$ caused by a lower \mej. 
(However, the initial high heating rate of \mtwo{}, which has a low \xlan{} and no appreciable fission or \al-decay, bucks this trend, reflecting the dependence of radioactivity on a complex interplay of factors. 
We caution that no single heuristic can account for the full variety of nucleosynthesis outcomes.)

\subsection{Ionization structure and light-curve evolution}
\label{subsec:doublepk}
 The effective heating for some of our models is higher than in many earlier studies of kilonovae \citep[e.g.][]{Barnes_2013,Barnes_etal_2016,Kasen.ea_2017Natur_gw.170817.knova.theory}, and the resulting hotter temperatures impact the ionization structure of the ejecta. 
The evolution of the ionization structure---and the increase in opacity that occurs when doubly-ionized lanthanide species recombine---in some cases leaves an imprint on the kilonova emission.

Where bound-bound absorption dominates opacity, opacity is sensitive to temperature,
peaking at ionization threshold temperatures and falling to local minima above these transition points.
This is because bound-bound opacity depends strongly on the number of distinct bound-bound transitions (``lines'') a photon encounters.
For a system in local thermodynamic equilibrium (LTE; a reasonable assumption for kilonovae during the photospheric phase), atoms in a given ionization state populate a broader distribution of energy levels as temperature increases from the ionization threshold.
At temperatures just above the threshold, atoms are confined to a handful of low-lying energy states, and there are comparatively few distinct lines contributing to the opacity \citep{Kasen_2013_AS,Tanaka_Hotok_rpOps,Tanaka.ea_2019_mnras_system.rproc.opacs}.

Kilonova opacity is almost entirely from lanthanide and actinide ions, and therefore varies more strongly with their ionization state than with that of lower-opacity species.
To illustrate this relationship, we calculate as a function of temperature the average lanthanide ionization state, \chilan,\footnote{
While \texttt{PRISM} predicts compositions containing both lanthanides and actinides, our radiation transport models include only lanthanides, due to a lack of atomic data (\S\ref{subsec:sedona_mods}). The addition of actinides is not expected to alter the relationships described above.}
and the mean opacity for conditions typical of our ejecta model ($\rho = 10^{-14}$ g$\,$cm$^{-3}$ and $t = 5$ days).
We use the Rosseland mean, $\bar{\kappa}_{\rm R}$, and average over a wavelength-dependent opacity that includes free-free and bound-free absorption, electron-scattering, and bound-bound transitions.
The latter we treat using the expansion opacity formalism \citep{Karp_1977, Eastman_1993}.

\begin{figure}
    \includegraphics[width=\columnwidth]{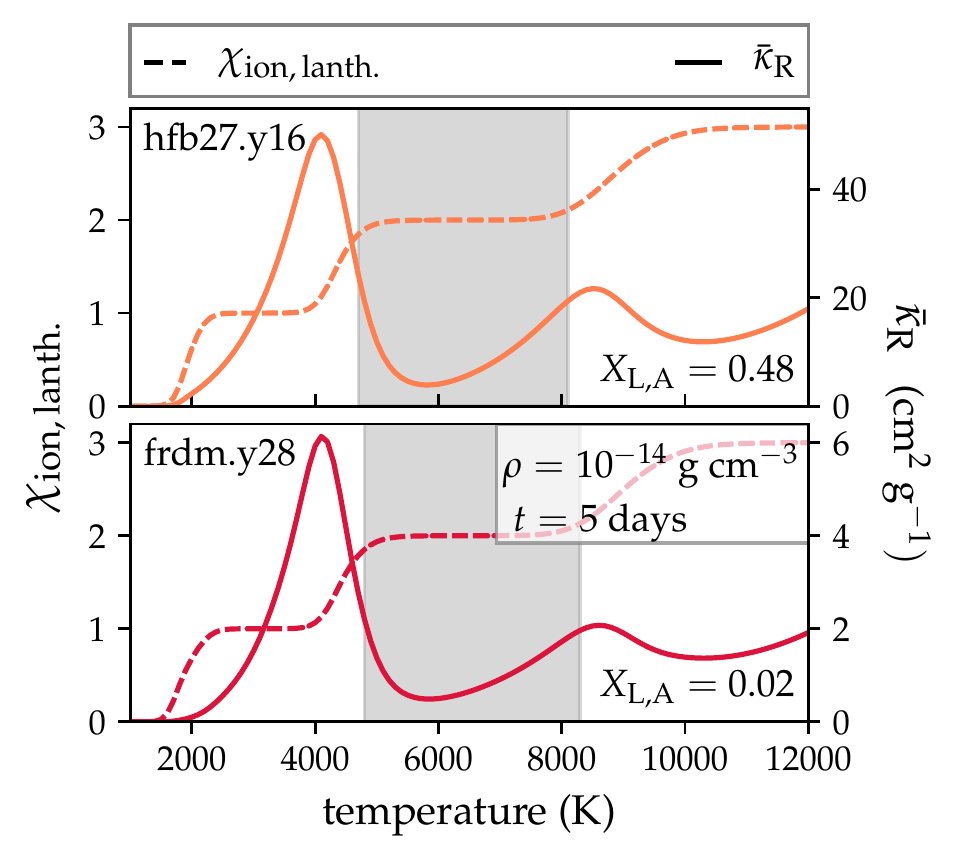}
    \caption{The average lanthanide ionization state, \chilan, and the Rosseland mean opacity, \kross, as a function of temperature, at density $\rho = 10^{-14}$ g$\,$cm$^{-3}$ and time $t = 5$ days after merger. The top (bottom) panel shows results for \mfour{} (\mtwo). Even for a low-lanthanide composition like \mtwo, the lanthanide ionization state determines \kross. Opacity is highest as the lanthanides transition from singly to doubly ionized, and passes through a local minimum just after.
    We have highlighted with a gray bar the temperature range corresponding to $1.8 \leq \chilan \leq 2.1$, which brackets this local minimum.}
    \label{fig:xion_kav}
\end{figure}

The results for \mtwo{} and \mfour{} (models 2 and 4, respectively) are presented in Figure~\ref{fig:xion_kav}.
Although the two models have the highest and lowest \xlan{} of our subset ($\xlan{4} = 0.48$, while $\xlan{2} = 0.02$),  the evolution of \kross\ tracks $\chilan$ in both cases.
(However, \xlan{} does affect the \emph{magnitude} of $\bar{\kappa}_{\rm R}$---note the different $y$-axis scales for each subplot.)
The opacity attains a maximum just as the lanthanide elements shift from singly to doubly ionized, but falls significantly once the transition is complete.

Where lanthanides are nearly exactly doubly ionized, a low-opacity ``window'' opens up in the ejecta.
Whether this window affects the observed luminosity depends on its location in the ejecta.
In many models of kilonovae, particularly those with lower \eeff, the region where $\chilan=2$ lies deep within the ejecta after very early times.
The effective opacity, and therefore the diffusion timescale that controls the evolution of the light curve, is then set by the higher opacity of the singly-ionized material outside the low-opacity region.
On the other hand, if a higher \eeff\ sustains until later times a spatially extended region where $\chilan =2$, the increase in opacity that occurs as the gas cools and lanthanide ions recombine to $\chilan=1$ \emph{can} influence the effective opacity of the system as a whole.

While it is a simplification to describe optically-thick, multi-wavelength radiation transport in terms of any single parameter, we find the opacity at the surface where the Rosseland mean optical depth, $\tau_{\rm R}(r) \equiv -\int_\infty^r \rho \bar{\kappa}_{\rm R} \mathrm{d} r'$, is unity to be a useful proxy for the effective opacity of the ejecta.
Specifically, the opacity at $\tau_{\rm R} = 1$ is correlated with the timescale on which the light curve evolves; an abrupt reduction in this opacity accelerates the light-curve evolution, while a sudden increase will slow it.

This can be understood if the $\tau_{\rm R} = 1$ surface is interpreted as a (rough) boundary separating an outer region, where radiation escapes easily, from an inner region where it diffuses out more slowly.
A rising opacity necessarily moves this boundary outward, forcing a larger fraction on the thermalized energy to diffuse out more slowly, while falling opacity has the opposite effect.

\begin{figure*}\includegraphics[width=\textwidth]{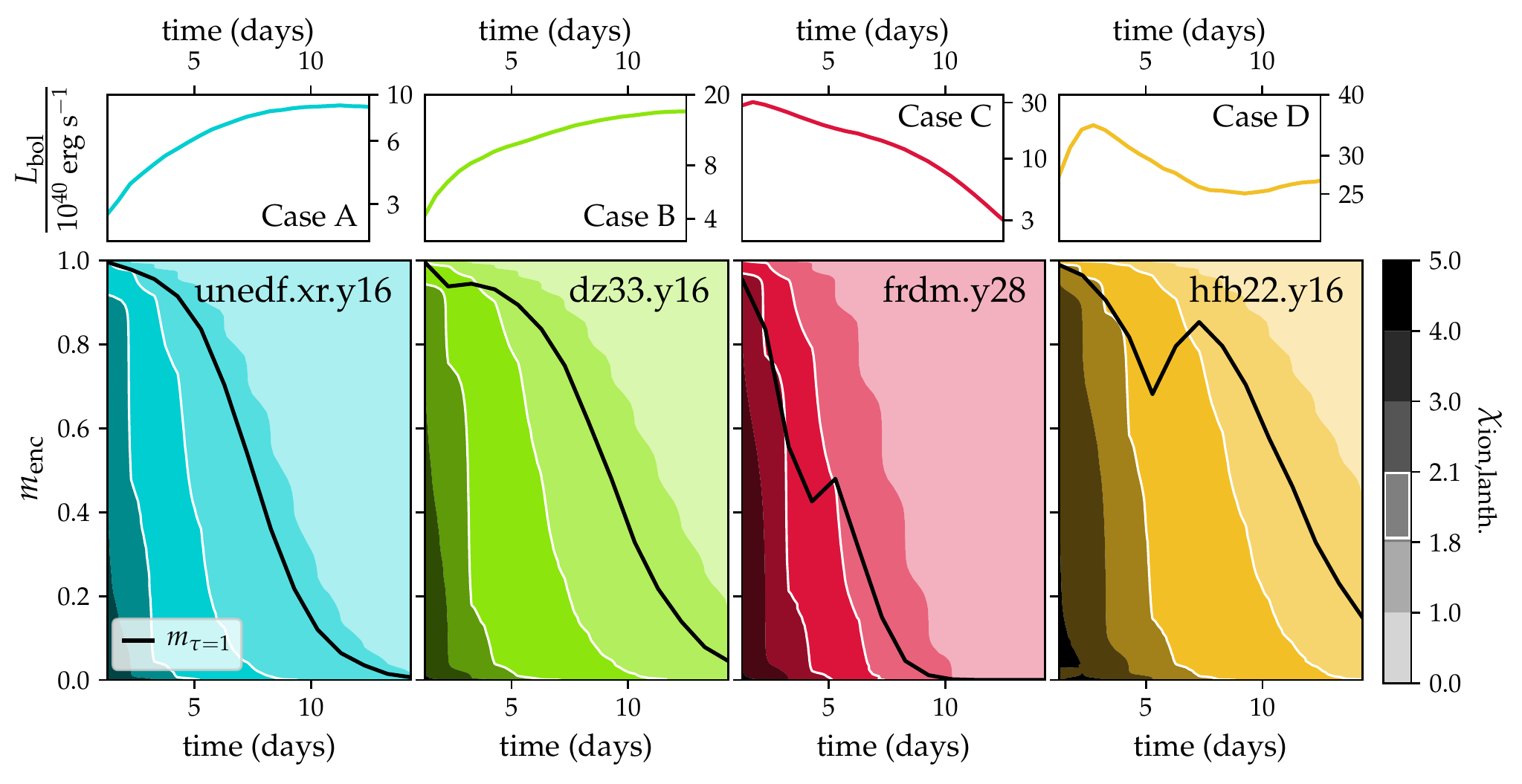}
    \caption{\textit{Top row:} Bolometric light curves.
    \textit{Bottom row:} The ionization state of lanthanides (\chilan; contours) and the location of the $\tau_{\rm R} = 1$ surface  as a function of time.
    Position in the ejecta is parameterized by the normalized interior mass coordinate.
    The region $1.8 \leq \chilan \leq 2.1$, which contains the local opacity minimum, is outlined in white for emphasis. 
    We have selected four models that represent Cases A--D, described above.
    \emph{First panel:} The $\chilan = 1$ region sets the opacity for the entirety of the kilonova. The low-opacity window does not effect the light curve.
    \emph{Second panel:} The $\tau_{\rm R} = 1$ surface intersects the low-opacity region early and only briefly, producing a minor excess at $t < t_{\rm pk}$.
    \emph{Third panel:} The effective opacity is set by $\chilan = 2$ through the light-curve peak. Recombination post-peak leads to a more gradual decline.
    \emph{Fourth panel:} The effective opacity is set by $\chilan=2$ during an early peak. Recombination significantly increases the effective opacity, creating conditions that can support a second rise to peak.
    }
    \label{fig:ion_phot}
\end{figure*}

Within our model subset, we identify four distinct ways that ionization state impacts light curves.
These are summarized in Figure~\ref{fig:ion_phot}, which maps out \chilan\ and the position of the $\tau_{\rm R} = 1$ surface as functions of the normalized interior mass coordinate $m_{\rm enc}$, defined as the fraction of \mej\ that lies interior to a given ejecta layer.
The bolometric light curves for the models presented have been re-plotted for convenience.
For an analytic discussion of how changing opacity influences light curves, see Appendix~\ref{appx:doublepk}.

\vspace{0.1mm}
\noindent\textbf{Case A: Standard evolution.} In the standard case, which is exemplified by \msev\ in Fig.~\ref{fig:ion_phot}'s first panel, sufficient cooling and a fairly high \xlan{} allow material with $\chilan =1$ to set the effective opacity over essentially the entire light curve.
The $\tau_{\rm R} = 1$ surface may skirt the $\chilan = 2$ region at very early times, when the entire ejecta is dense and multiply ionized, but at all times $\gtrsim$ a few hours, it tracks the singly-ionized ejecta layer.
In addition to \msev, \msix, \meig, \mten, and \melev\ belong to this category. All (see Fig.~\ref{fig:heff_lbol}) have low \eeff\ and $\xlan{} \geq 0.12.$

\vspace{1mm}
\noindent\textbf{Case B: Minor early excess.} In models with slightly greater \eeff{}, represented by \mfive{} in  Fig.~\ref{fig:ion_phot} (panel 2), higher temperatures produce a doubly ionized layer near the ejecta's edge, reducing the system's effective opacity and allowing the light curve to rise sharply.
However, \eeff{} is not large enough to maintain this state for long, and the opacity increases as lanthanides recombine, helped along by a high \xlan{} that exacerbates the opacity difference between singly and doubly ionized gas.
The $\tau_{\rm R} = 1$ surface quickly jumps to the singly ionized region, signaling an increase in effective opacity that flattens the light-curve rise and creates the appearance of early excess luminosity.
In addition to \mfive, early excesses are seen for \mone{} and especially \mfour{}.
All have higher \eeff\ and greater \xlan{} ($0.22 \leq \xlan{} \leq 0.48$) than the models of Case A.

\vspace{1mm}
\noindent\textbf{Case C: Extended decline phase.}
In the third case, a low \xlan{}, perhaps assisted by a large \eeff{}, keeps the effective opacity low through and beyond the light curve peak, even as the ejecta's outer layers start to recombine.
This is illustrated by \mtwo{} in the third panel of Fig.~\ref{fig:ion_phot}.
The lower opacity of the $\chilan = 2$ region causes the luminosity to peak sharply and fall off swiftly.
However, the boundary where $\chilan
$ transitions from 2 to 1 recedes inward with time, and eventually there is enough singly-ionized ejecta to drive the opacity up.
The increasing opacity, which manifests in the abrupt displacement of the $\tau_{\rm R} = 1$ surface, slows the light curve's decline, producing a ``shoulder'' feature in the light-curve tail.
(An analogous effect is responsible for the infrared rebrightening of Type Ia supernovae \citep{Kasen_2006.ApJ_SNIa.ir.2max}.)
This is seen for \mnine{} as well as \mtwo.

\begin{figure*}[t]
    \includegraphics[width=\textwidth]{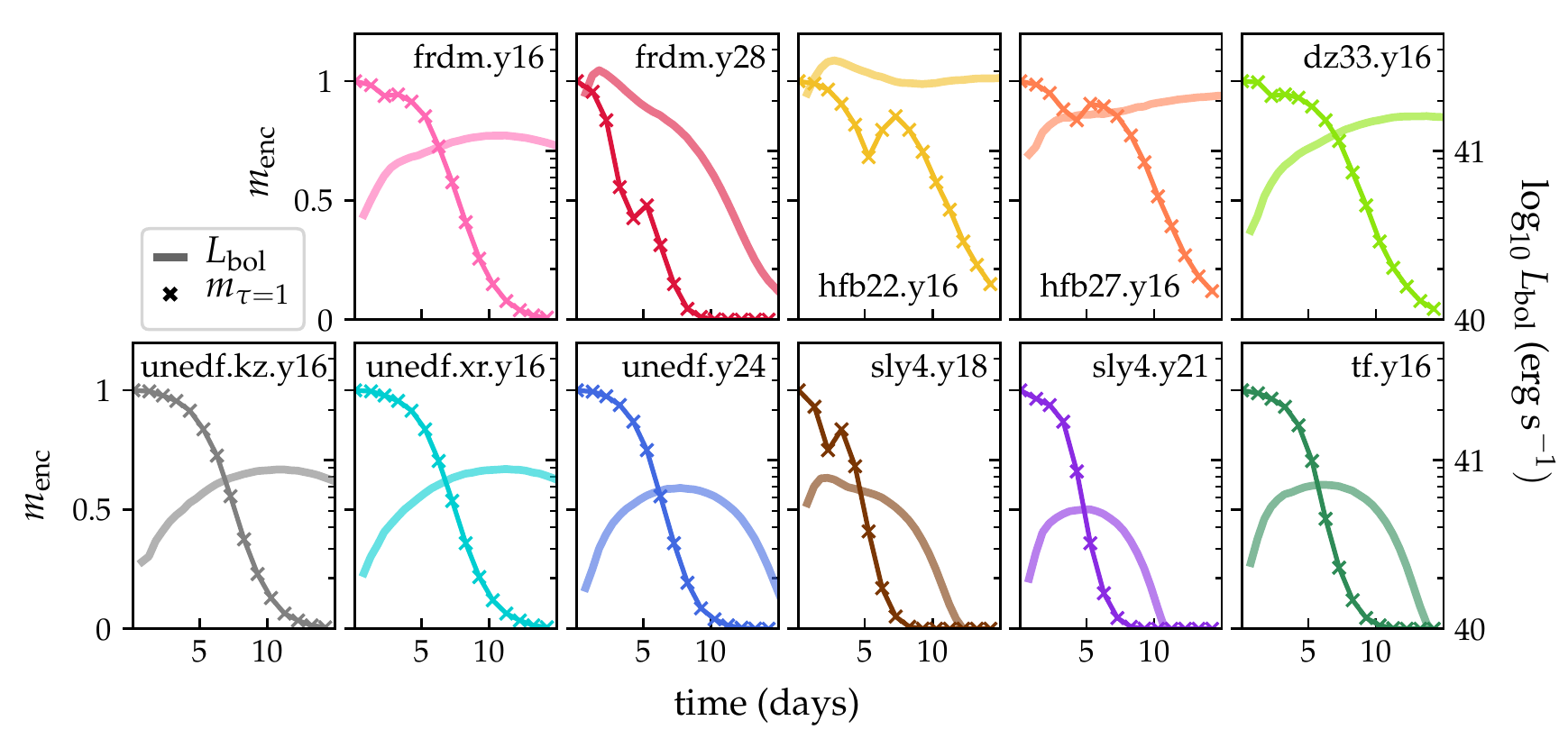}
    \caption{Bolometric light curves (light-colored lines) and the location of the $\tau_{\rm R} = 1$ surface (x's) for all models.
    Increases in $m_{\tau=1}$ indicate a rising effective opacity, driven by recombination, which affects the light curves.}
    \label{fig:lc_mphot}
\end{figure*}

\vspace{1mm}
\noindent\textbf{Case D: Dual-phase light curve.}
Under certain conditions, recombination allows a light curve to reach two distinct bolometric peaks. 
This is the case for \mthree, shown in the fourth panel of Fig.~\ref{fig:ion_phot} as the sole exemplar of Case D. 

Early on, strong heating produces an extended highly ionized region, which sets a low initial effective opacity that supports a rapid light-curve rise.
As indicated by the $\tau_{\rm R} = 1$ surface, this lower opacity is preserved through the first bolometric peak at $t \approx 3$ days.

While the opacity of the doubly ionized region is low relative to singly ionized material, it is still fairly high in an absolute sense because of the high \xlan{} $({=}0.37)$ of \mthree{} (model 3).
However, early peaks are possible even for high-opacity systems if \eeff{} drops sharply (\S\ref{subsec:lc_eeff}), and the decline of \eeff{,3} around the time of the first peak is particularly steep ($\rm{d}\log\eeff{,3}/\rm{d}\log{t} \approx -2$ for $1 \leq t/{\rm day} \leq 5$; see Fig.~\ref{fig:heff_lbol}).

The interplay of \xlan{} and \eeff{} can be understood by contrasting the light curves of the two HFB models.
Both have high \eeff{}, but compared to \mthree{}, \mfour{} has a slightly shallower early decline and a higher \xlan{} = 0.48.
Thus, while \mthree{} forms a peak, \mfour{} merely exhibits a pronounced early excess.
The situation changes if \xlan{} is reduced.
This can be seen in the right-hand panel of Fig.~\ref{fig:lcs_xe_std}, which shows light curves calculated using our numerically determined \eeff{} and a standard composition with $\xlan{}=0.15$. 
The brightest light curve in that panel corresponds to \mthree{}, and the second-brightest to \mfour{}.
The reduced opacity for this variant of \mfour{} allows an early peak to form, despite the fact that \eeff{} declines less steeply for \mfour{} than for \mthree{}.
As these examples show, an early peak depends on \eeff{} \emph{and} opacity; 
the higher \xlan{} is, the steeper the \eeff{} an early peak requires.

The second peak of \mthree{} is due to an increase in its opacity that coincides with a dramatic flattening of its \eeff{}.
As can be inferred from Fig.~\ref{fig:ion_phot}, for \mthree{}, the opacity change from recombination is significant; it pushes the $\tau_{\rm R} = 1$ surface out almost to the edge of the ejecta, effectively resetting the ejecta to a condition similar to that of the pre-peak phases of Case A (standard) kilonovae.

At the same time, the heating rate becomes much shallower (${\rm d}\ln \eeff{,3}/{\rm d}\ln t \rightarrow -0.5$).
This is the result of the growing importance of the long-lived fissioning isotopes \iso{Cf}{254} and \iso{Rf}{269} (Fig.~\ref{fig:fiss_iso}), which increases thermalization (Fig.~\ref{fig:ft_tot}) and softens the decline of \eeff{}.
The extra energy injection from the flatter \eeff{} replenishes the internal energy, providing power for a second peak at $t_{\rm pk,2nd} \approx 16$ days.

\vspace{1 mm}
As these four cases demonstrate, the effect of high heating rates and highly-ionized regions in the ejecta exist on a continuum.
While more than half our models are impacted in at least a minor way, the strongest effects are observed only for \mthree{} and \mfour{}, which stand out for their high \xlan{} and extremal \eeff{}.

To illustrate the range of outcomes, we show in Figure~\ref{fig:lc_mphot} the position of the $\tau_{\rm R} = 1$ surface and the bolometric luminosity for all models.
Increases in effective opacity, which appear in the plots as abrupt increases in the mass coordinate where $\tau_{\rm R} = 1$, are indeed correlated with apparent discontinuities in the light curves.

\subsection{Spectral Energy Distribution}

\begin{figure*}
\includegraphics{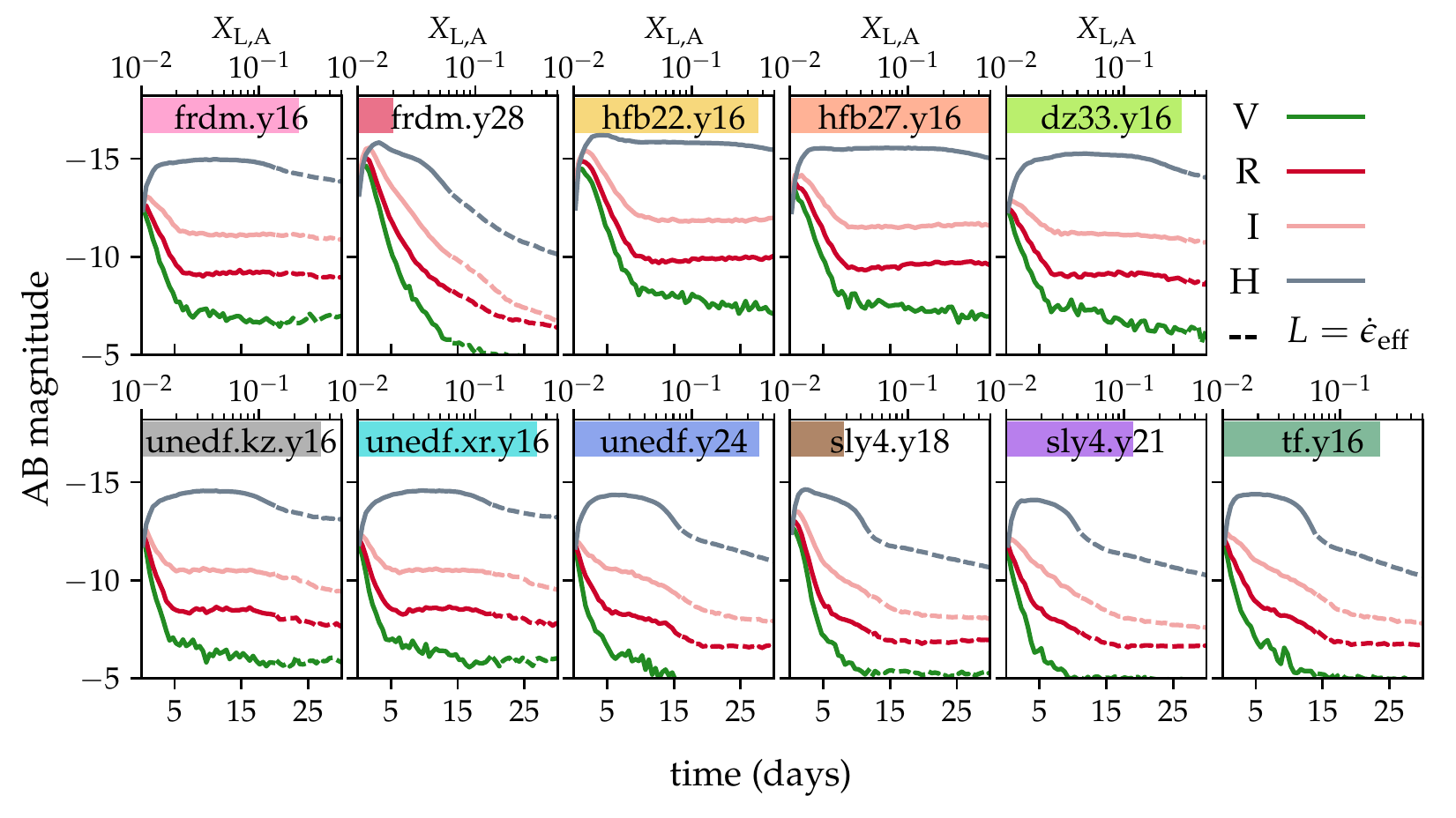}
    \caption{Absolute AB magnitudes in Bessel $V$, $R$, $I$, and $H$ bands as a function of time, with \xlan{} for each model indicated by the shaded bar at the top of each panel.
    For all models, optical emission drops off quickly, and much of the energy comes out in the NIR. 
    Dashed lines indicate times after which bolometric luminosity and the effective heating rates have converged.
    At these times, the system is optically thin, and the assumption of LTE may introduce errors into synthetic spectra.}
    \label{fig:broadband}
\end{figure*}

Select optical and near infrared (NIR) broad-band light curves for each of our models are presented in Figure~\ref{fig:broadband}. 
There is some variation in the magnitude and morphology of the light curves, which parallels variability in bolometric luminosity.
However, the evolution of the spectral energy distribution (SED) is overall consistent model-to-model.
Emission at optical wavelengths is suppressed after very early times, with much of the radiation emerging instead in the NIR.
This behavior conforms to expectations established by earlier theoretical studies of lanthanide-rich kilonovae \citep{Barnes_2013, Tanaka_Hotok_rpOps}, and to observations of the red kilonova of GW170817 \citep[][and references therein]{Aracavi.ea_2017Natur_gw170817.lco.emcp.disc,Drout.ea_2017Sci_gw.170817.emcp.disc,Villar.ea_2017ApJ_gw.170817.knova.agg}.

While the distinct spectra of kilonovae that are rich ($\xlan{} \gtrsim 10^{-2}$) v. poor ($\xlan{} \lesssim 10^{-3}$) in lanthanides and actinides has been well-documented \citep[e.g.,][]{Metzger.Fernandez_2014_red.blue.bhform.knova,Tanaka.ea_2019_mnras_system.rproc.opacs}, the effect on emission colors of very high \xlan{}, like those considered here, has received less attention.
Consistent with earlier studies, we find that, generally speaking, lower \xlan{} is correlated with bluer colors. 
However, this correlation weakens with increasing \xlan{} and is further complicated by the effects---for some models---of an early low-opacity phase associated with a ${\sim}$doubly ionized ejecta.
As a result, there is not necessarily an obvious observational fingerprint that can distinguish between moderately and extremely lanthanide-rich compositions, at least if uncertainties in \eeff{} are taken into account.

\begin{figure}\includegraphics[width=\columnwidth]{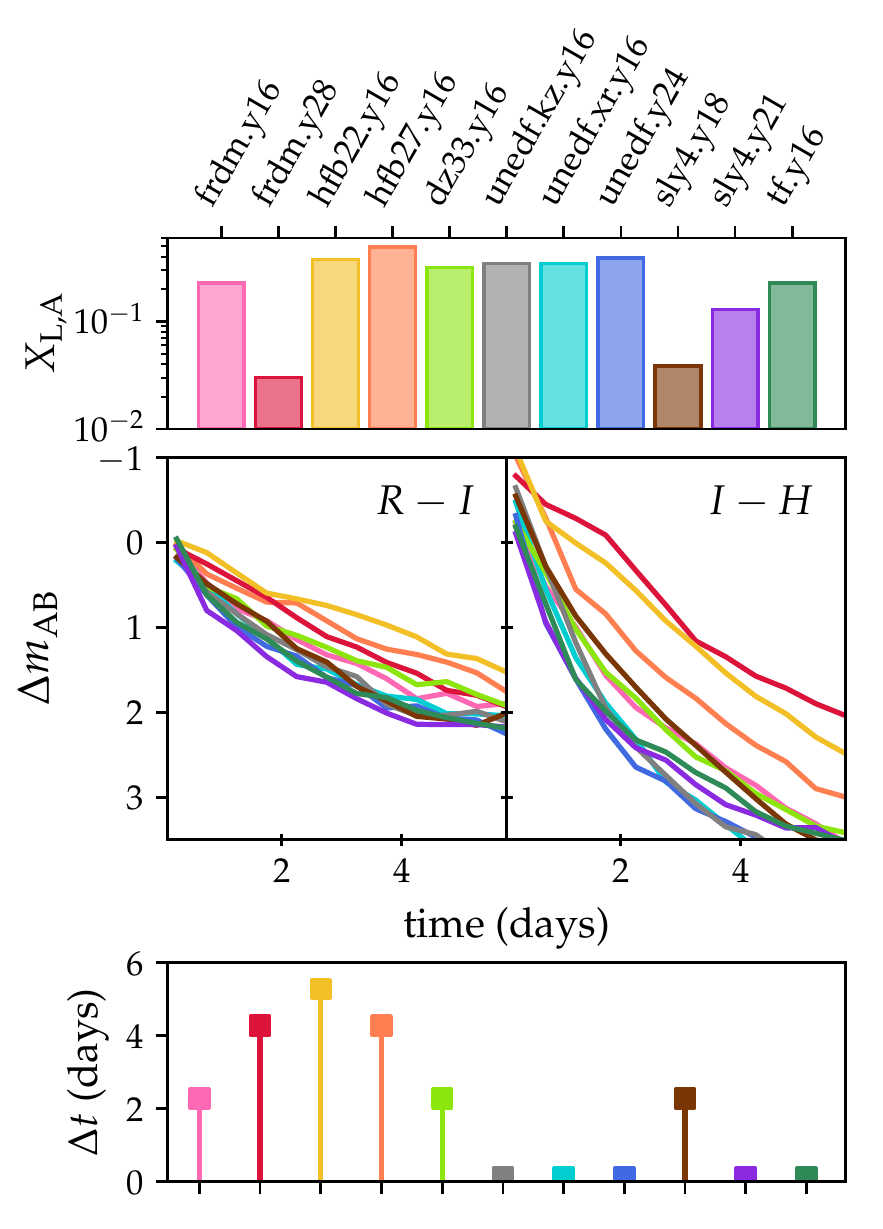}
    \caption{The colors at early times are not a straightforward function of \xlan{}.
    \textit{Top panel:} \xlan{} for all models.
    \textit{Middle panels:}
    Select colors for all models at $t \leq 5$ days. The models with the bluest colors do not necessarily have the lowest \xlan{}.
    \textit{Bottom panel:}
    An estimate of the length of time each model's photosphere spends in the hot $\chilan \gtrsim 2$ region.
    The low opacities that allow the photosphere to form at higher temperatures may be caused by low \xlan{} or by enhanced ionization from high \eeff{}.}
    \label{fig:early_color}
\end{figure}

The assumption that low \xlan{} leads to bluer colors is least reliable at early times, when even lanthanide-rich outflows may shine blue if
high levels of ionization lower the effective opacity and allow the photosphere to form deep in the ejecta where temperatures are hotter.
This is shown in Figure ~\ref{fig:early_color}, which presents $R-I$ and $I-H$ colors for all models at $t \leq 5$ days.
We also show \xlan{} for each model in the top panel, while the bottom panel estimates how long each model's photosphere, defined as the surface where $\tau_{\rm R} = 1$, remains in the $\chilan=2$ region.
The bluest models at this epoch include those with the lowest \xlan{} (e.g. \mtwo{}, with \xlan{} = 0.02), and those with some of the highest (\mthree{}, with \xlan{} = 0.37, and \mfour{}, with \xlan{} = 0.48).
In the latter case, ionization by large \eeff{} compensates for the high \xlan{}, lowering the opacity enough to place the photosphere at temperatures hotter than the first ionization threshold of lanthanides.

\begin{figure}
    \includegraphics[width=\columnwidth]{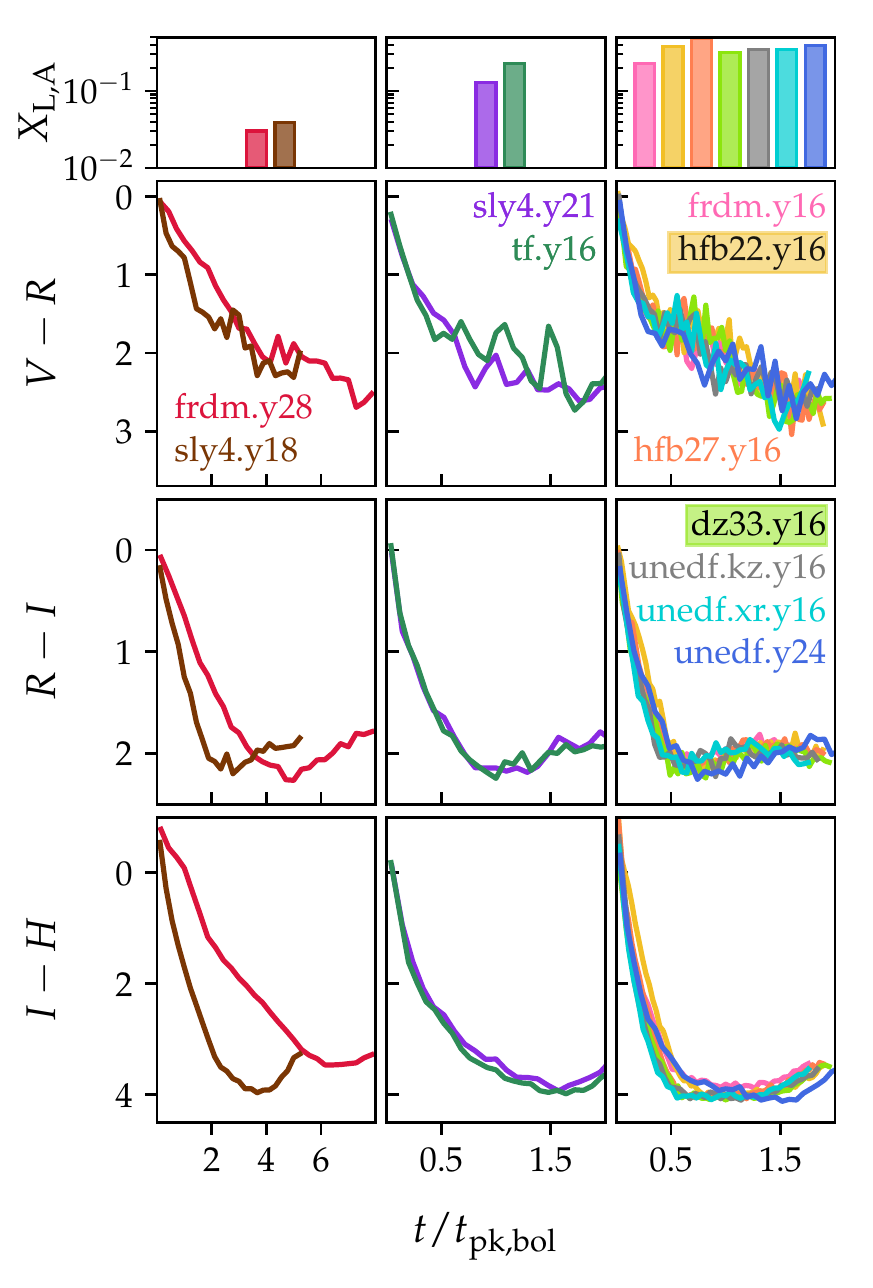}
    \caption{Color evolution of the model suite as a function of $t/t_{\rm pk}$.
    For \mthree{}, we have normalized to the second bolometric peak, $t_{\rm pk,2nd}$ = 15.75 days.
    Models with different \eeff\ and \xlan{} (plotted in the panels of the top row) nonetheless show fairly consistent color evolution.}
    \label{fig:clrev_tnorm}
\end{figure}

This early phase is transient though, and for nearly all models, the photosphere forms at $\chilan=1$ for most of the kilonova's duration. 
Because of this, the kilonova has fairly consistent colors at a given phase in its evolution. 
This is illustrated in Figure~\ref{fig:clrev_tnorm}, which shows $V-R$, $R-I$, and $I-H$ colors for all models, as a function of time normalized to bolometric peak time. 
For \mthree, we have normalized to the second bolometric peak, $t_{\rm pk, 2nd} = 15.75$ days.

The models \mtwo{} (\xlan{} = 0.02) and \mnine{} (\xlan{} = 0.03) appear as outliers in the left-hand column of Fig.~\ref{fig:clrev_tnorm}.
The low opacity of these models produces an early light-curve peak and keeps the photosphere in the $\chilan=2$ region through maximum light.
As a result, their emission stays bluer out to later $t/t_{\rm pk}$.

The differences among the remaining models are much subtler. 
Model \mten{}, shown in the middle column of Fig.~\ref{fig:clrev_tnorm}, has an intermediate \xlan{} ($=0.12$) and slightly bluer colors at $t \leq t_{\rm pk}$ than the higher-\xlan{} models of the right-hand column.
However, its color evolution is consistent with that of \melev{} (middle column), despite \melev{} having $\xlan{}=0.22$, nearly twice that of \mten{}.
Conversely, \mone{} has \xlan{} equal to that of \melev{}, but shares the marginally redder colors of the models in the right-hand column, which span $0.22 \leq \xlan{} \leq 0.48$.

Despite this variance in \xlan{}, the models of the right-hand column have $R-I$ ($I-H$) colors at bolometric peak concentrated in the range 2.0 -- 2.2 (3.8 -- 4.1).
The effects of \xlan{} on color are stronger for low \xlan{}; the increase in \xlan{} from \mnine{} ($\xlan{} = 0.03$) to \mten{} ($\xlan{} = 0.12$) raises $R-I$ ($I-H$) at peak by 0.9 (${\gtrsim}1.6$).
This mainly reflects the formation of the photosphere in the $\chilan = 1$ region for \mten{}, v. the $\chilan = 2$ region for \mnine{}.
Broadly similar behavior over a wide range of high \xlan{} underscores the challenge of using color as a diagnostic of composition in the lanthanide-rich regime.

The root of this similarity is
the strong dependence of opacity on temperature, and the diminishing marginal impact of increasing \xlan{}.
As has been pointed out elsewhere \citep[][see also Fig.~\ref{fig:xion_kav}]{Barnes_2013, Tanaka_Hotok_rpOps}, the sharp decrease in opacity as lanthanides and actinides transition from singly-ionized to neutral
ensures that, under most conditions, the photosphere forms over a narrow range of temperatures around the first lanthanide ionization threshold.
Changes to the opacity can nudge the photospheric temperature one way or the other, and the emerging pseudo-blackbody spectrum is modified by line-blanketing and absorption features.
However, the effect of \xlan{} on the opacity, and thus on kilonova SEDs, is strongest for low \xlan{} \citep{Kasen_2013_AS}.
Once the lanthanide concentration reaches some threshold, the impact of increasing \xlan{} is minimal, because the opacity has been saturated. 
Strong lines are already absorbing so effectively that additional strength does not meaningfully increase their optical depth.

\begin{figure}
    \includegraphics[width=\columnwidth]{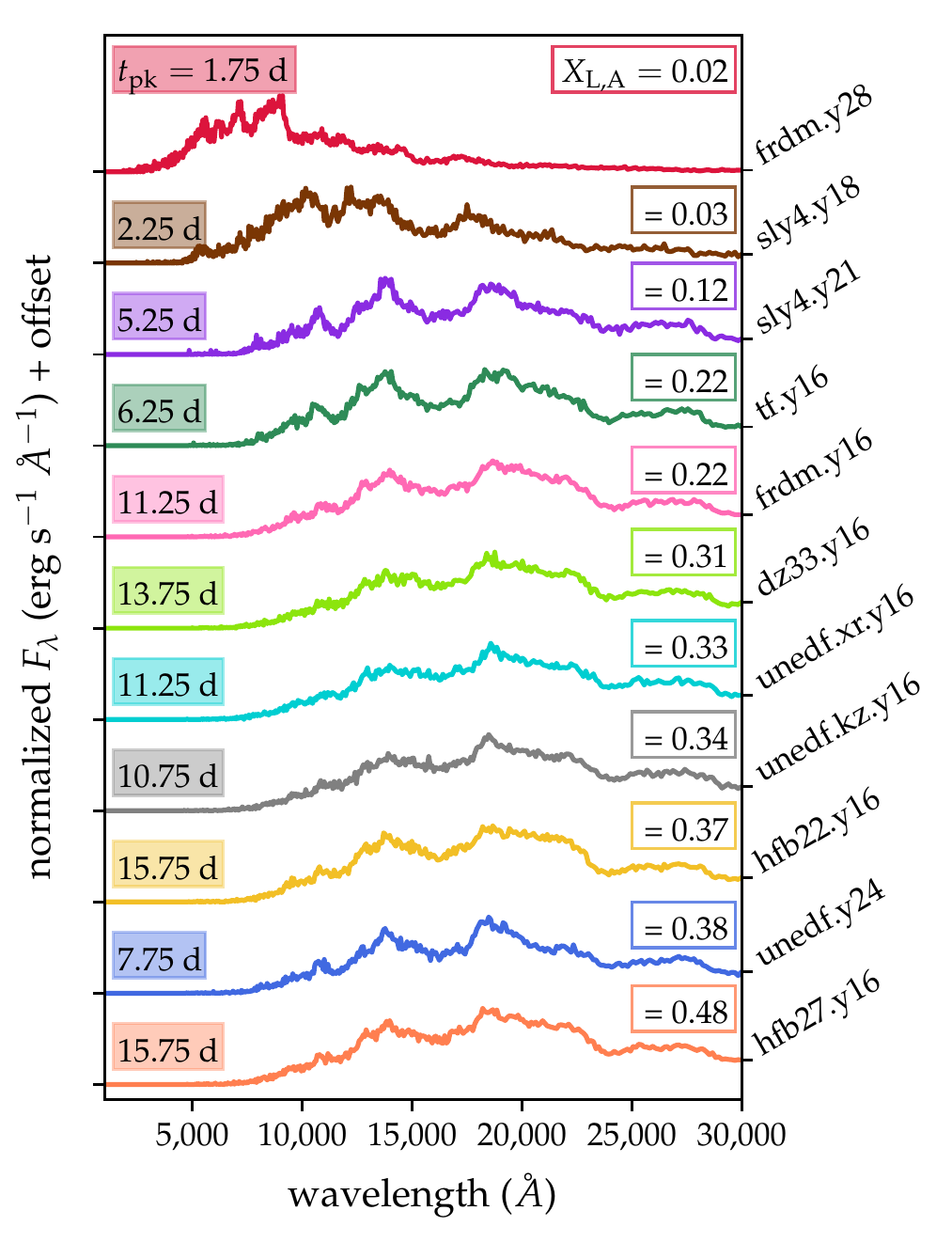}
    \caption{The spectrum at bolometric peak for all our models, with $t_{\rm pk}$ and \xlan{} noted for each. We have again adopted $t_{\rm pk} = 15.75$ days for \mthree. 
    Once \xlan{} surpasses some threshold 
    $X_{\rm L,A} \sim 0.1$, the effects of increasing \xlan{} on the peak spectrum are small.}
    \label{fig:peak_spectrum}
\end{figure}

To illustrate this, we plot in Figure~\ref{fig:peak_spectrum} the spectrum at bolometric peak for each of our models. 
Models \mtwo{} (2) and \mnine{} (9) have the bluest peak spectra, which is not surprising given their low opacity (\xlan{2} = 0.02 and \xlan{9} = 0.03) and the fact that their photospheric emission at peak originates from a hotter, more highly-ionized ejecta layer.
The other models, despite the large range of lanthanide concentrations represented ($0.12 \leq \xlan{} \leq 0.48$) show only minor spectral variation.
(And even \mtwo{} and \mnine{} evolve redward with time.)

Of course, the spectrum depends not only on \xlan{}, but also on how \xlan{} is distributed among individual lanthanides and actinides. 
While a detailed accounting of the lines most important for spectral formation in kilonovae is a long-term project, theoretical work \citep{Kasen.ea_2017Natur_gw.170817.knova.theory,Even.ea_2020.apj_comp.effects.kne} suggests that certain elements may have more influence than others.
Similarities in spectra (observed or simulated) may then point to comparable abundances of these dominant species, rather than identical total \xlan{}.
Both opacity saturation and the potential spectral dominance of few species suggest that a large range in \xlan{} can nonetheless produce spectra with similar shapes.

\section{Discussion and Conclusion}\label{sec:conc}
 
Using a set of calculations that self-consistently accounts for absolute heating by radioactivity, thermalization of decay energy, and radiation transport of thermal photons in the kilonova ejecta, we have demonstrated that varying nuclear physical inputs in \rp{} simulations produces significant diversity in predicted kilonova emission.
Our calculations, which are restricted to a single choice of ejecta mass (\mej) and velocity (\vej), produce kilonovae with a broad range of peak luminosities and timescales. 

Our analysis revealed that effective heating \eeff\ strongly influences kilonova rise times and light-curve shapes, complicating the well-established relationship between time-to-peak and opacity (and therefore lanthanide and actinide content, \xlan{}).
We find that high \eeff{} has a significant impact on the ionization structure of the ejecta, which in turn influences its effective opacity.
We articulate four distinct ways ionization state affects the kilonova evolution, and even identify circumstances that enable unexpected outcomes, such as a kilonova with two bolometric maxima, or a lanthanide-rich outflow that powers an early, surprisingly blue peak.
Finally, we examined the evolution of our model kilonovae's spectral energy distributions, and showed that the high \xlan{} values of some of our models have only a minor effect on the emerging spectrum at various stages of the kilonova evolution.
This indicates the difficulty of establishing an upper (v. a lower) limit on \xlan{} based on spectral observations alone.

To understand the implications of our findings for interpreting observations of kilonovae, it is useful to review some key determinants of kilonova properties.
Our work has focused on how the choice of nuclear parameters influences the predicted kilonova associated with a single ejecta model.
It is now instructive to invert the problem and consider how adjusting \mej\ and \vej\ would affect kilonovae for a fixed set of nuclear physics parameters.
As alluded to in \S\ref{subsec:lc_eeff} (Eq.~\ref{eq:tpeak}), decreasing \mej\ and/or increasing \vej\ reduces the optical depth of the ejecta, and encourages more rapid light-curve evolution. 
However, since thermalization depends on local mass density, it is also sensitive to \mej\ and especially to \vej\ ($\rho \propto \mej \vej^{-3}$; see \citetalias{Barnes_etal_2016}).
Both the heating near peak and the late-time decline of \eeff\ are thus subject to change with the ejecta configuration. 

This dependence, along with the range of absolute heating rates and opacity/\xlan{} in our model subset, suggest that distinct sets of nuclear physics parameters could nonetheless produce kilonovae with very similar light curves if \mej\ and \vej\ are varied.

We defer a full exploration of the inverted problem to future work, but discuss here briefly how the interlocking \mej-\vej-\eeff{}-\xlan{} degeneracy identified in this work may be addressed.
Spectroscopic observations of future kilonovae will be particularly valuable, both because the width of absorption features provides an independent measure of ejecta velocity, and because spectra offer the possibility of pinpointing features associated with particular atoms or ions \citep[e.g.][]{Smartt.ea_2017Natur_gw170817.empc.disc,Watson.ea_2019Natur_gw.170817.knova.strontium}.
The first constrains ejecta parameters, and thus the permitted nuclear physics inputs.
The second may allow certain sets of nuclear physics inputs to be ruled out based on discrepancies between predicted and observed abundances or abundance ratios.
Kilonovae with extreme luminosities may also exclude some nuclear physics parameters, e.g. by requiring unphysically high \mej.

Nuclear physics experiments will also be crucial. Measurements of the properties of exotic nuclei, 
like those ongoing at RIKEN in Japan \citep[e.g.,][]{Wu+2020} and the CARIBU facility at Argonne National Lab \citep[e.g.,][]{Orford+2018}, and planned for the TRIUMF Advanced Rare Isotope Laboratory (ARIEL), the FAIR accelerator facility at GSI, and the Facility for Rare Isotope Beams \citep[FRIB;][]{FRIB2018}, will reduce nuclear physics uncertainties and confine plausible mass models to a more narrow region of parameter space. 
One of the goals of this investigation was to motivate experimental measurements of the nuclei with the greatest potential to promote this narrowing; see \S 5 of \citetalias{Zhu.ea_kntherm_in.prep} for a detailed inventory of key nuclei and reactions.

This work has taken an important step toward delineating the kilonova diversity allowed by current uncertainties in nuclear physics.
Our results suggest a few clear avenues for further investigation.
First, single-trajectory models, like those employed here as a proof-of-principle, are idealizations.
True kilonova outflows will be comprised of gas elements characterized by a range of astrophysical variables ($s_{\rm B}$, $\tau_{\rm exp}$, and \ye{}).
A natural next step is to explore the range of heating such composite models permit.

In a similar vein, it will be worthwhile to examine how constraining final abundances to approximately match solar values would affect our results.
Ideally, models with close-to-solar abundance yields will be constructed from multiple trajectories to more accurately reflect realistic merger conditions.

Such models will offer a more precise accounting of the \mej-\vej-\eeff{}-\xlan{} degeneracy identified here.
They will be complemented by a more data-rich description of nuclear physics in the heavy, neutron-rich regime, enabled by pioneering experiments at facilities like RIKEN, CARIBU, ARIEL, and FRIB and---one hopes---by a growing catalog of mergers and their electromagnetic (EM) counterparts discovered by gravitational-wave observatories like LIGO/Virgo and their network of EM observing partners.

These efforts will provide the multi-messenger community with the tools required to interpret kilonova emission with unprecedented accuracy, and therefore better understand merger-driven mass ejection and the chemical enrichment of the Universe by merging compact binaries.

\vspace{\baselineskip}
\noindent\textbf{Acknowledgements: }
J.B. acknowledges support from the National Aeronautics and Space Administration (NASA) through the Einstein Fellowship Program, grant number PF7-180162, and through grant NNX17AK43G; as well as from the National Science Foundation (grant number AST-2002577). 
The work of Y-L.Z., K.L., N.V., G.C.M., M.R.M., T.M.S., and R.S. was partly supported by the Fission In R-process Elements (FIRE) topical collaboration in nuclear theory, funded by the U.S. Department of Energy. 
Additional support was provided by the U.S. Department of Energy through contract numbers DE-FG02-02ER41216 (G.C.M), DE-FG02-95-ER40934 (R.S. and T.M.S.), and DE-SC0018232 (SciDAC TEAMS collaboration, R.S. and T.M.S). 
R.S. and G.C.M also acknowledge support by the National Science Foundation Hub (N3AS) Grant No. PHY-1630782. M.R.M. was partially supported by the Laboratory Directed Research and Development program of Los Alamos National Laboratory under project number 20190021DR. 
Los Alamos National Laboratory is operated by Triad National Security, LLC, for the National Nuclear Security Administration of U.S. Department of
Energy (Contract No. 89233218CNA000001). 
This work was partially enabled by the National Science Foundation under Grant No. PHY-1430152 (JINA Center for the Evolution of the Elements).
The work of K.L. was supported partially through EUSTIPEN (Europe-U.S. Theory Institute for Physics with Exotic Nuclei), which is supported by FRIB Theory Alliance under DOE grant number DE-SC0013617.

\newpage

\appendix
\counterwithin{figure}{section}
\section{Complete Thermalization Efficiency Results}\label{appx:ft_all}
\begin{figure}[b]\includegraphics[width=\textwidth]{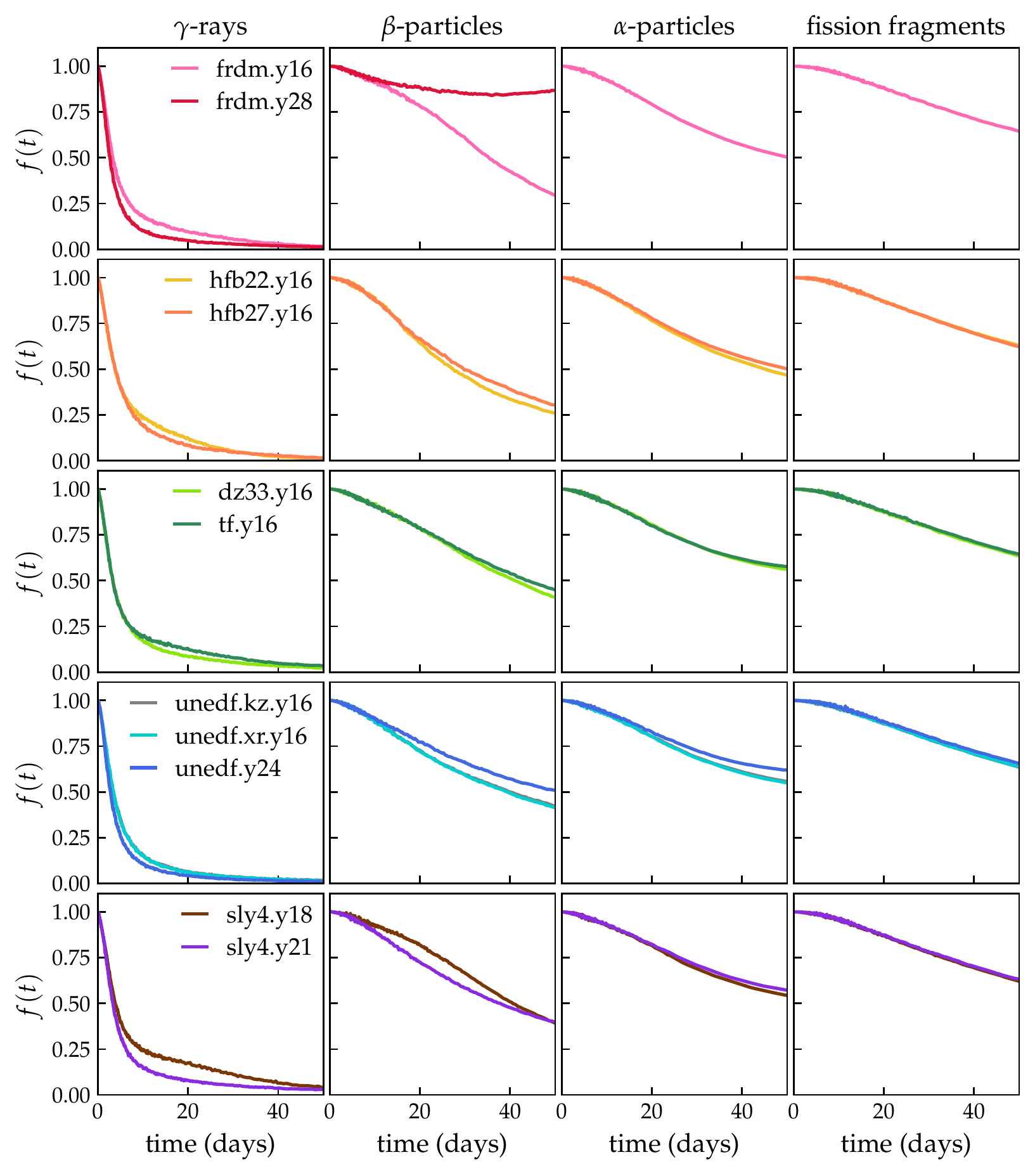}
\caption{Thermalization efficiencies of all particles, for all models, as a function of time. 
The \al-decay and fission for \mtwo{} (model 2) were minimal, and $f_{\al,2}(t)$ and $f_{\rm ff,2}(t)$ were not calculated.}
\label{fig:allft_appx}
\end{figure}

In an elaboration of Fig.~\ref{fig:ft_all}, we present the time-dependent thermalization efficiencies $f_{\rm i}(t)$ of all particles for all models in 
Figure~\ref{fig:allft_appx}.

\section{Analytic models of double-peaked light curves
}\label{appx:doublepk}

In the following, we use simple analytic models of radioactively powered transients to explore the conditions that produce a light curve with two distinct bolometric peaks.
Guided by the discussion of \S\ref{subsec:doublepk}, we assume that increasing opacity is required for such behavior.

\subsection{Derivation of the model}\label{appdxb:model}

We start with the equation for conservation of energy in a simple one-zone, homologously expanding ejecta with a time-varying gray opacity, $\kappa(t)$,
\begin{align}
    \frac{\mathrm{d}E_{\rm int}}{\mathrm{d}t} &= \dot{Q} - \frac{E_{\rm int}}{t} - L \nonumber \\
                                              &= \dot{Q} - \frac{E_{\rm int}}{t} - \frac{4\pi v c}{3 M \kappa(t)} E_{\rm int} t, \label{eq:dedt_a} 
\end{align}
where \eint\ is the ejecta's internal energy, $\dot{Q}$ is the heating from radioactivity (analogous to \eeff{} in previous sections), and $L$ is the luminosity.
In the second line, which makes use of the diffusion equation, $v$ is the ejecta velocity, $M$ is the ejecta mass, and $\kappa(t)$ is a time-dependent opacity.

We assume that opacity starts at some initial value \ki\ and asymptotically approaches a value $\kf > \ki$.
For simplicity, we express opacity as $\kappa(t) = \kf k(t)$ and define a light-curve timescale \tlc{} in terms of \kf: $\tlc^2 \equiv 3M\kf/4 \pi v c$ \citep[e.g.,][]{Arnett_1982_Sne}.
This allows us to rewrite $L$ as
\begin{align}
    L =  \frac{\eint t}{\tlc^2 k(t)}\label{eq:L_redef}.
\end{align}

Taking the derivative of $L$ with respect to time provides a new expression for the time derivative of \eint.
Accounting for the time-dependence of $k$, $L$ evolves as
\begin{align}
    \ddf{L}{t}& = \frac{1}{\tlc^2}\left[\ddf{\eint}{t}\frac{t}{k} + \frac{\eint}{k} -\frac{\eint t}{k^2}\ddf{k}{t} \right] \\
    \rightarrow \ddf{\eint}{t} &= \frac{\tlc^2 k}{t}\ddf{L}{t} - \frac{\eint}{t}\left(1 - \frac{t}{k}\ddf{k}{t}\right).
\end{align}
When substituted into Eq.~\ref{eq:dedt_a}, this yields, after some rearranging,
\begin{align}
    \frac{\tlc^2 k}{t}\ddf{L}{t} + \frac{\eint}{k}\ddf{k}{t} + L &= \dot{Q} \nonumber \\
    \ddf{L}{t} + L\left( \frac{1}{k}\ddf{k}{t} + \frac{1}{\tlc^2}\frac{t}{k}\right) = \frac{\dot{Q}t}{\tlc^2 k}. \label{eq:dldt}
\end{align}

\begin{figure}
\centering
\includegraphics[]{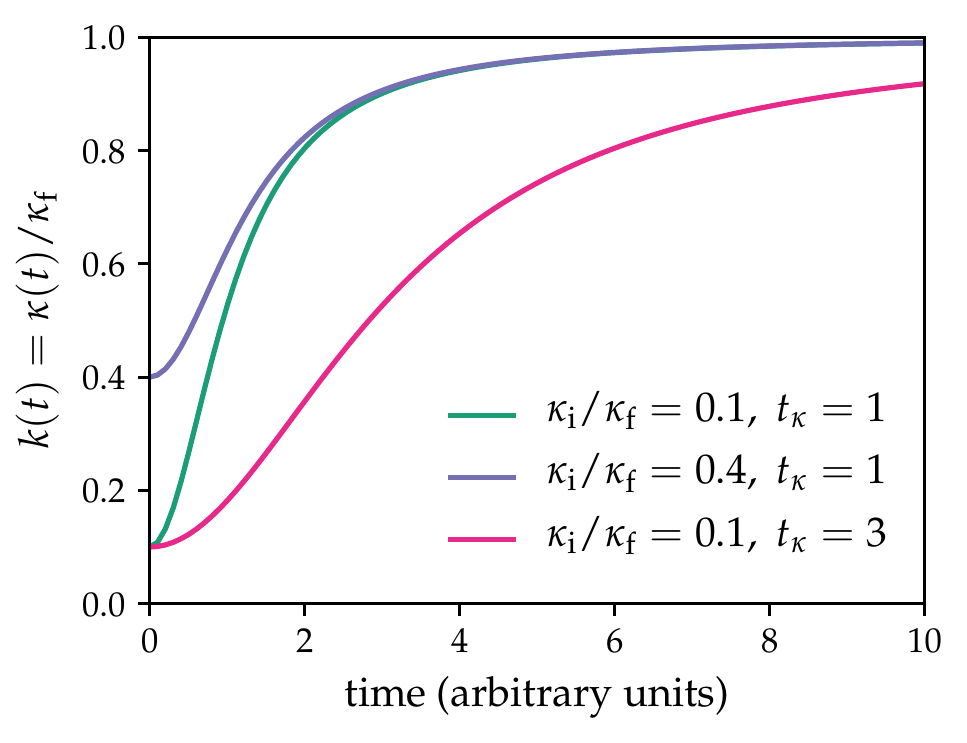}
\caption{The behavior of the function $k(t)$ (Eq.~\ref{eq:koft}) which is responsible for the time-dependence of the opacity. The value of $k(t)$ increases asymptotically from $k = \ki/\kf$ to $k \rightarrow 1$. The parameter $b$ is related to \ki\ and \kf\ by $b = \ki/(\kf-\ki)$ The timescale of the transition is set by \tk.  }
\label{fig:appdx_kform}
\end{figure}

Eq.~\ref{eq:dldt} is an ordinary linear differential equation, and can be solved with an integrating factor $u(t)$, once the form of $k(t)$ is known. 
(This is the motivation for making $k$ a function of $t$ directly, rather than \eint.)
We choose
\begin{align}
    k(t) = \frac{t^2 + b\tk^2}{t^2 + (1+b) \tk^2},\label{eq:koft}
\end{align}
which has the desired limiting behavior.
In this formulation, the constant $b$ is determined by the requirement that $k(t=0) = \ki/\kf$, while $t_{\kappa}$ sets the timescale of the transition from \ki\ to \kf.
We show $k(t)$ for a few sets of parameters $b$ and $t_{\rm k}$ in Fig.~\ref{fig:appdx_kform}.

With this choice, the second term in parentheses in the L.H.S. of  Eq.~\ref{eq:dldt} becomes
\begin{align*}
    \frac{1}{\tlc^2}\frac{t}{k} = \frac{1}{\tlc^2}\left[t + \frac{t}{(t/\tk)^2 + b}\right].
\end{align*}
This allows the integrating factor to be expressed analytically,

\begin{align}
    u(t) &= \exp[\smallint P(t)\; \dd{t}] \nonumber \\
     P   &= \frac{1}{k}\ddf{k}{t} + \frac{1}{\tlc^2}\left[t + \frac{t}{(t/\tk)^2 + b} \right] \\
     \rightarrow u(t) &= k(t) (t^2 + b \tk^2)^{(\tk^2/2\tlc^2)} \exp[ t^2/2\tlc^2]\label{eq:integ_factor}
\end{align}

We can now write an expression for the time-dependent luminosity,
\begin{align}
    \begin{split}
    L(t) = k(t)^{-1} \, (t^2 + b\tk^2)^{(-\tk^2/2\tlc^2)} \, \exp[-t^2/2\tlc^2]
    \times \left[ \vphantom{\int\limits_1^2} L_0  (\ki/\kf)(b \tk^2)^{(\tk^2/2\tlc^2)} \: +  \right. \\
    \left. \int\limits_0^t \frac{\dot{Q}(t')t'}{\tlc^2} \, (t'^2 + t^2_\kappa b)^{(\tk^2/2\tlc^2)} \, \exp[t'^2/2 \tlc^2] \, \dd{t'} \right]
    \end{split}\label{eq:Lt_nonorm}
\end{align}

Eq.~\ref{eq:Lt_nonorm} becomes slightly more palatable if expressed in terms of dimensionless units.
Defining $\tau = t/\tlc$ and $\tauk = \tk/\tlc$, we arrive at a final expression for luminosity,
\begin{align}
\begin{split}
  L(\tau) = k(\tau)^{-1} \, (\tau^2 + b\tauk^2)^{(-\tauk^2/2)} \, \exp [-\tau^2/2] \: \times \:
  \left[ \vphantom{\int\limits_0^t} L_0(\ki/\kf) (b\tauk)^{(\tauk^2/2)} + \right. \\
  \left. \int\limits_0^\tau \dot{Q}(\tau')\tau' (\tau' + \tauk^2 b)^{(\tauk^2/2)} \: \exp[\tau'^2/2] \; \dd{\tau '} \right]
\end{split}\label{eq:Lt_norm}
\end{align}
In practice, the initial luminosity $L_0$ will be negligible compared to the luminosity at $t > 0$, once radioactive heating has begun, and the approximation $L_0 = 0$ can simplify Eq.~\ref{eq:Lt_norm} further.

\subsection{Model light curves}

The light curves described by Eq.~\ref{eq:Lt_norm} will depend on the form of $\dot{Q}$, as well as \tk\ and the ratio \ki/\kf. We start by considering simple power-law heating, $\dot{Q} \propto \tau^{-1}$, to illustrate the effects of the opacity parameters.
Since our interest is in the shape of the light curves, we do not normalize \qdot, and the luminosity scales in our models are therefore not physically meaningful. 
In all calculations in this section, we have assumed the contributions of radioactive heating dominate the system's initial luminosity, and set $L_0 = 0$.

\begin{figure}
    \centering
    \includegraphics{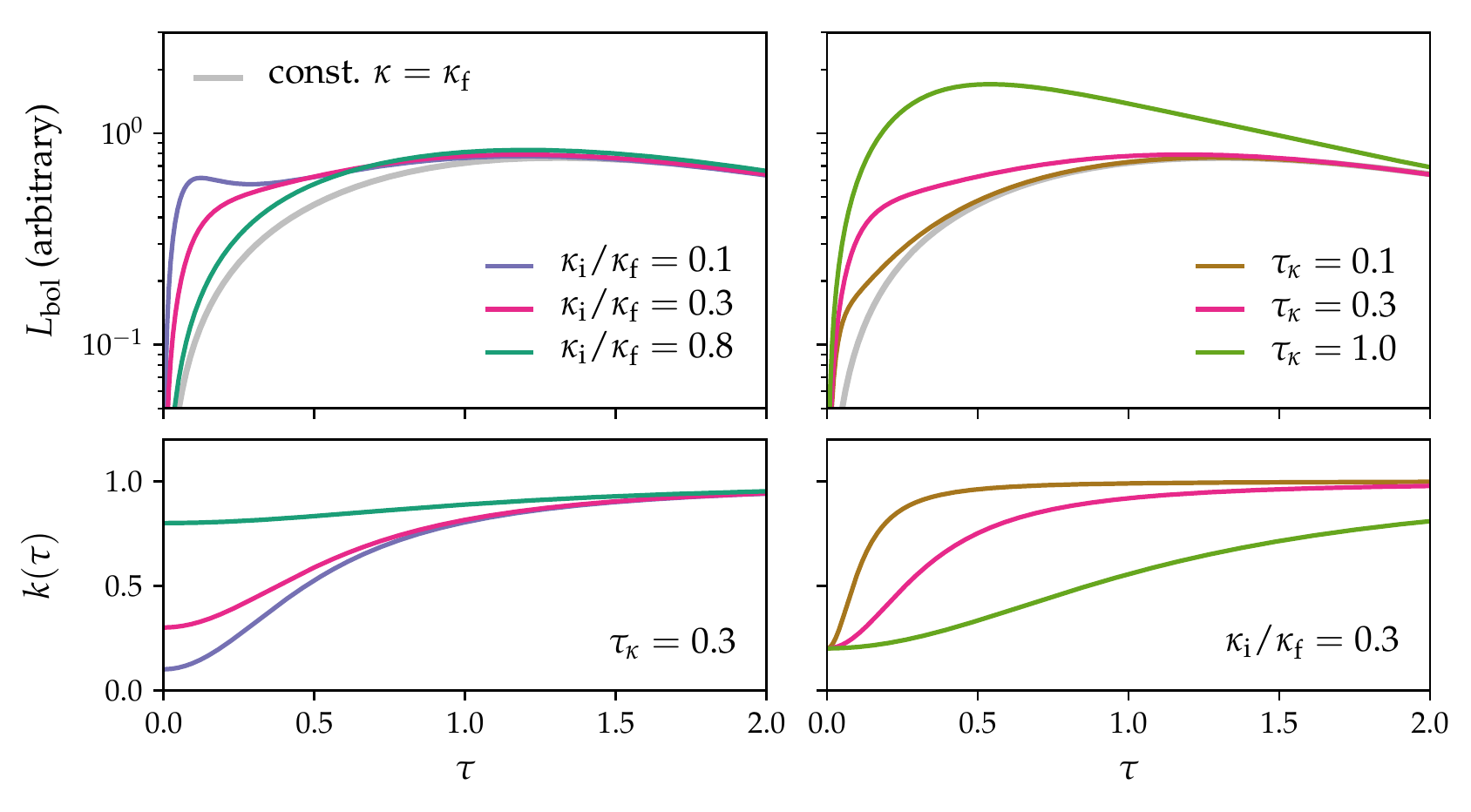}
    \caption{The effect of $k(\tau)$ (\textit{bottom panels}) on analytic light curves (\textit{top panels}).
    In all cases, we have set $\dot{Q} \propto \tau^{-1}$.
    Gray curves in the top panels correspond to constant $\kappa = \kf$.
    \textit{Left panels:} We vary \ki/\kf{} and fix $\tk = 0.3$. A lower \ki\ leads to brighter rises.
    \textit{Right panels:} We set $\ki/\kf{} = 0.3$, while \tauk\ varies. For $\tauk < 1$, the increasing opacity at $\tau \sim \tauk$ slows the light-curve rise.}
    \label{fig:lcmods_q1}
\end{figure}

In Figure~\ref{fig:lcmods_q1}, we show the effect on light curves of \ki/\kf{} (left column) and the timescale \tauk\ (right column).
Light curves are shown in the top row, while the associated $k(\tau)$ are plotted below. 
For comparison, we show as a gray line the light curve produced for a constant $k(t) = 1$ (i.e., $\kappa(\tau) = \kf$).
As can be seen in the left-hand panels, the impact on light curves of an early low opacity is enhanced luminosity during the light curve rise. 
The smaller \ki/\kf{} is, the more pronounced the enhancement is, and the more obvious the effect of the increasing opacity at $\tau \sim \tauk = 0.3$. 
From the model with $\ki/\kf = 0.1$, we see that double-peaked light curves are possible for this choice of \qdot, but require a fairly extreme opacity increase.
All models in Fig.~\ref{fig:lcmods_q1} peak  globally at $\tau \approx 1$.

The models in the right-hand column tell a similar story.
The larger \tk\ is, the longer the system maintains its early low opacity and, as a result, the brighter the light-curve rise is relative to the case with constant $\kappa=\kf$.
When the opacity transition occurs prior to peak, the new, longer diffusion time induced by the higher \kf\ reduces the slope of the light curve's rise.
We note that for $\tauk \gtrsim 1$, \ki\ is more characteristic of the light-curve opacity than \kf, and scaling to $\tlc$ (defined in terms of \kf) via the $\tau$ parameter becomes less logical.
(This is the reason the $\tauk = 1$ model in Fig.~\ref{fig:lcmods_q1}'s top-right panel appears to have an earlier, more luminous peak than the other light curves.)

As discussed above (\S\ref{sec:nucphys} and \S\ref{subsec:ft_net}), \rp{} heating across large swaths of the nuclear physics parameter space is \emph{not} well-represented by a single power-law.
We have also argued (\S\ref{subsec:doublepk}), that changes in the heating rate may support the formation of double-peaked light curves.
This motivates us to consider broken-power law forms of $\dot{Q}$,
\begin{align}
    \dot{Q}(\tau) &\propto \tau^{-\zeta}\nonumber \\
    \zeta &=
    \begin{cases}
        & \zeta_1, \:\: \tau \leq \tau_{\rm Q} \\ 
        & \zeta_2, \:\: \tau > \tau_{\rm Q}
    \end{cases} \nonumber
\end{align}

\begin{figure}
\centering
\includegraphics[width=\textwidth]{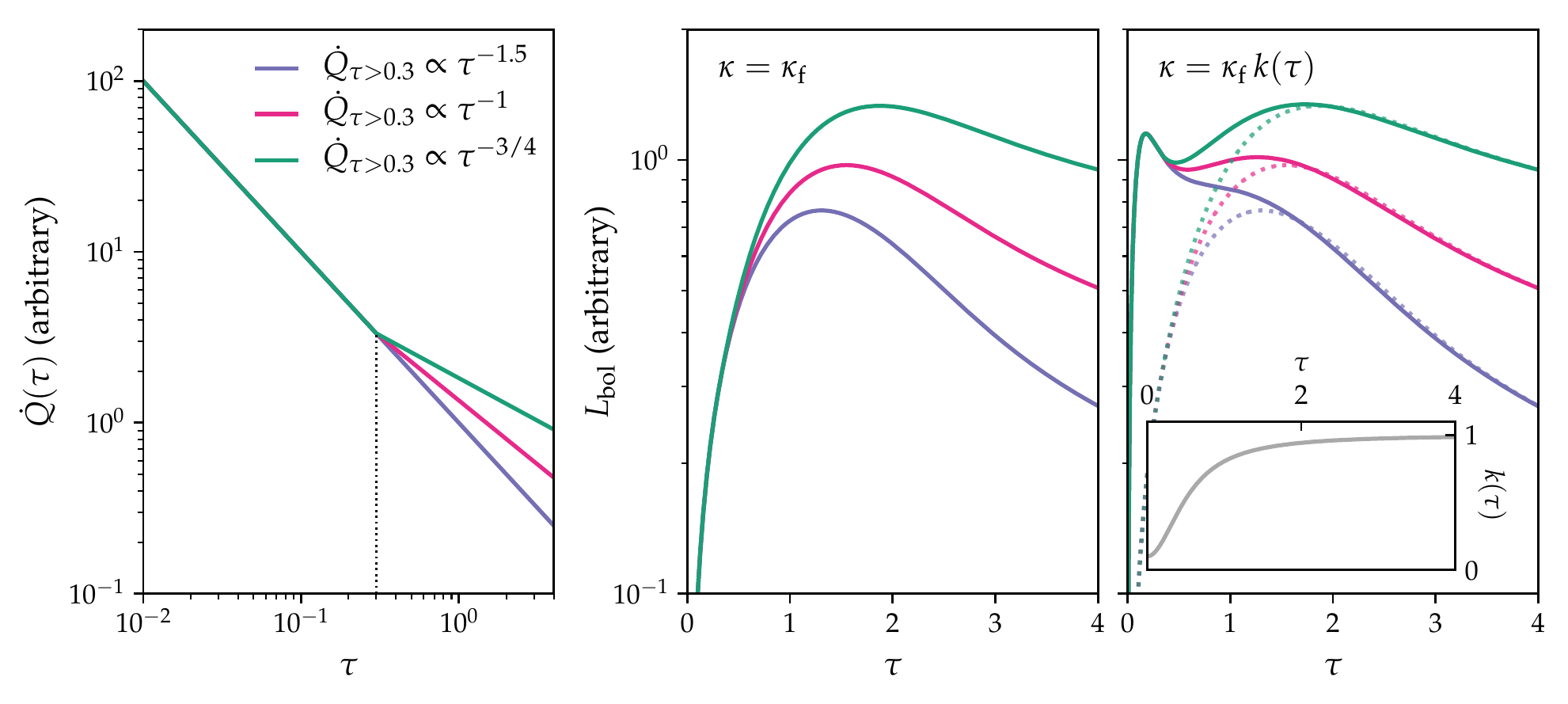}
\caption{Analytic light curves for a broken-power law \qdot. 
\textit{Right panel:} The \qdot's considered in this figure. All curves have $\dot{Q}(\tau \leq 0.3) \propto \tau^{-1}$. In two of the three models, the slope of \qdot\ increases at $\tauq=0.3$.
\textit{Middle panel:} Analytic light curves for a constant opacity. The more luminous teal and magenta curves reflect flatter \qdot\ at $\tau > 0.3$, but no early peak is formed. 
\textit{Right panel:} Varying-opacity light curves, with $\ki/\kf{} = 0.1$ and $\tauk = 0.5$. (We show $k(\tau)$ in the inset axes). All models form an early peak associated with the early low opacity, but the formation of a second peak---local or global---depends on the slope of \qdot\ at $\tau > \tauq$. The constant opacity light curves are plotted as dotted lines to demonstrate convergence.
}
\label{fig:plawq}
\end{figure}

Fig.~\ref{fig:plawq} shows model light curves for a handful of broken-power law $\dot{Q}$, all with $\tau_{\rm Q} = 0.3$ and $\zeta_1 = 1$, for constant and time-evolving opacity.
All models with a time-dependent $\kappa(\tau)$ have $\ki/\kf{} = 0.1$ and $\tauk = 0.5$.
As can be seen in the middle panel, when $\kappa$ is constant, changes to $\dot{Q}$ do not lead to any early peaks or other surprising features, though the flattening of $\dot{Q}$ does lead to brighter, later peaks.
The former is expected based on the extra energy supplied by a flatter \qdot, while the latter is due to a shallow \qdot's improved ability to compensate for energy lost to diffusion and adiabatic expansion (\S\ref{subsec:lc_eeff}).

When $\kappa$ increases in time, changes to \qdot{} strongly affect the  light curve, as shown in Fig.~\ref{fig:plawq}'s right-hand panel.
All models in this panel have an early peak, enabled by a low \ki/\kf.
However, whether that peak is the only peak produced, a global peak in a dual peaked light-curve, or a local peak followed by a global peak depends on how dramatically \qdot{} flattens. 
For $\zeta_2 = 1/2 \:\: (3/4)$, enough energy is injected into the system to support a second rise to a global (local) maximum. 
For $\zeta_2 = \zeta_1 = 1$, no secondary peak is present, and the rising opacity merely produces a ``shoulder'' feature, similar to that observed in some of our \texttt{Sedona} models (\S\ref{subsec:doublepk}).

Allowing \qdot\ to vary further expands the parameter space under consideration, and introduces a new timescale, $\tau_{\rm Q}$, into our analysis. 
Our model light curves now depend on the change in opacity, $\ki/\kf$; \tauk, the timescale of the opacity transition; the slope(s) of \qdot; and the timescale \tauq\ on which any changes to that slope occur.

\begin{figure}
    \centering
    \includegraphics[width=\textwidth]{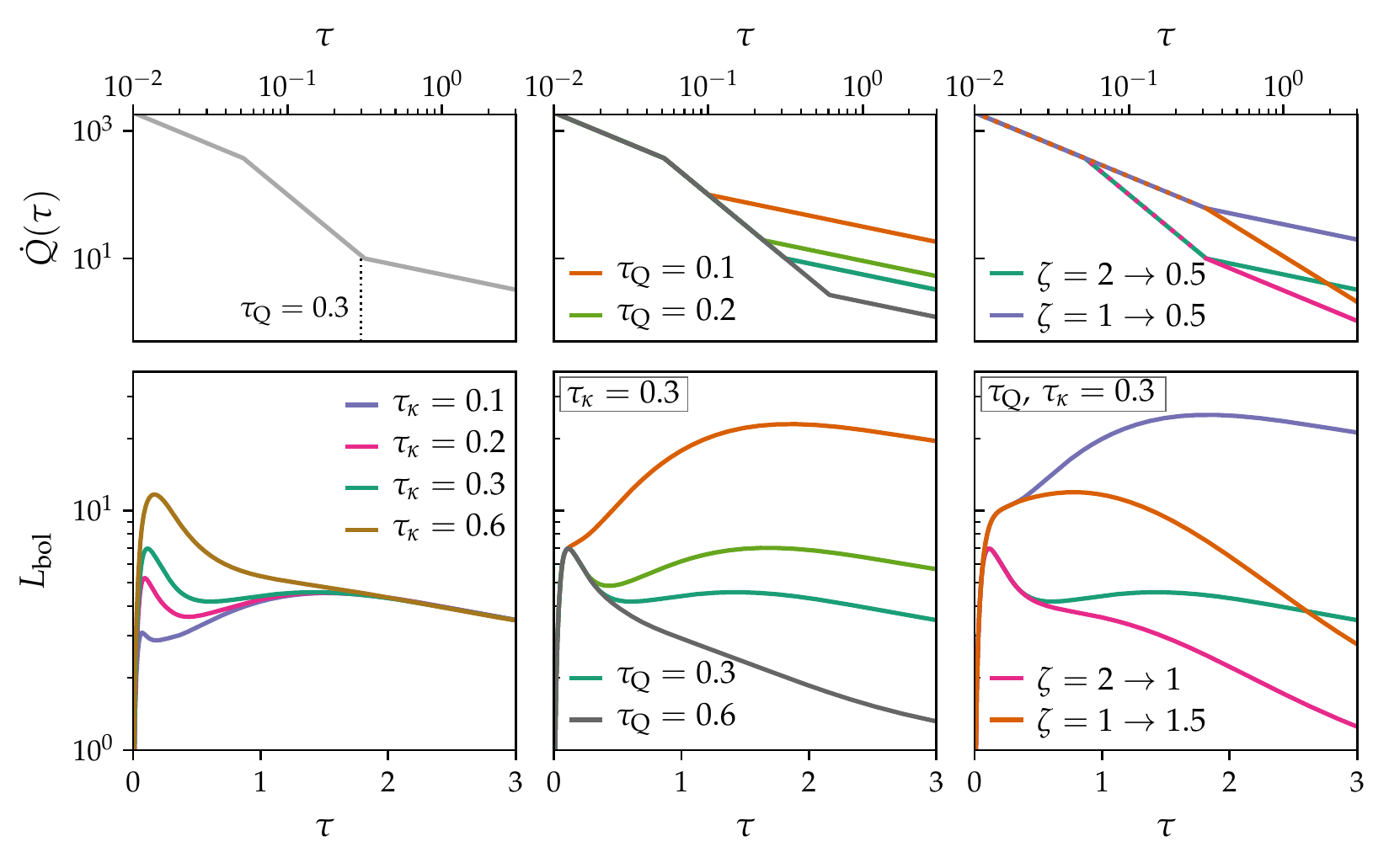}
    \caption{The effects of \tauk, \tauq, and the power-law indices of \qdot on light curves. The top panels show \qdot, while the lower panels present the associated light curves.
    \textit{Left panel:} \tauk is varied while the heating rate, defined by $(\zeta_1, \zeta_2, \tauq) = (2,0.5,0.3)$ is held constant.
    \textit{Middle panel:} We fix $\tauk = 0.3$, and ($\zeta_1, \zeta_2) = (2,0.5)$, but adjust \tauq.
    \textit{Right panel:} We vary the power-law indices $\zeta_1$ and $\zeta_2$ for $\tauq = \tauk = 0.3$.
    The form of \qdot\ as well as the choice of \tauq\ and \tauk, both in absolute terms and in relation to each other, effect the light curve.
    }
    \label{fig:appx_modlc_summ}
\end{figure}

While we do not attempt to map out the full parameter space described above, we explore a limited slice of it in Figure~\ref{fig:appx_modlc_summ}, in order to illustrate the most important trends and highlight how certain model parameters interact.
All of the models in Fig.~\ref{fig:appx_modlc_summ} have a broken-power law \qdot.
For most \qdot, we have set the first power-law index $\zeta_1 = 2$.
This is steeper than in Fig.~\ref{fig:plawq}, but is a good approximation to the effective heating rate of the model (\mthree) that produces a double-peaked light curve in our numerical calculations.
However, such a steep heating rate is not expected to characterize very early-time \rp{} decay (see Fig.~\ref{fig:model_qandy} of this work, or e.g., \citet{Korobkin_NSM_rp}, \citet{Lippuner.Roberts_2015_rproc.params.knova}, or \citet{Wu.Barnes.ea_2019_late.time.knova.rproc}).
To avoid overheating at early times as \qdot{} is extrapolated back toward $\tau = 0$, we force \qdot\ at $\tau < 0.05$ to decay no more steeply than $\tau^{-1}$.

The \qdot's for each model are plotted in the top panels while the corresponding light curves are presented below.
For all models, we have adopted a moderate opacity ratio $\ki/\kf{} = 0.2$, though (Fig.~\ref{fig:lcmods_q1}) a smaller value would allow double-peaked light curves in a larger region of the parameter space.
The left and middle columns show the effect of varying the timescales \tauk\ and \tauq.
In the left-hand column, $\tauq = 0.3$, while \tauk\ is a free parameter, while in the middle column, \tauk\ is set to 0.3, and we vary \tauq.

Both panels suggest that the appearance of two well-defined peaks depends on having $0.1 \lesssim \tauk \approx \tauq \lesssim 0.6$.
If either timescale is too low, the system does not spend enough time in the regime of steeply declining \qdot\ and/or low opacity for those early conditions to leave a strong imprint on the light curve, which simply evolves in accordance with $\kf$ and $\zeta_2$.
This is the case for the violet curve in the lower-left panel and the orange curve in the lower-middle panel.
On the other hand, if $\kappa$ or \qdot{} transitions too slowly, the light curve evolution, determined by \ki\ and $\zeta_1$, is largely complete by the time the shifts take place. 
In these instances, represented by the brown and gray curves in the lower-left and -middle panels, respectively, the increasing opacity and/or the flattening of \qdot\ impact the tail of the light curve, but do not lead to the formation of a new peak.

In the right-hand panels, we explore how the impact of $\zeta_1$ and $\zeta_2$ on light curves, while holding $\tauk = \tauq = 0.3$. 
While all models in the lower-right panel show some sort of early feature, the data in this panel suggest that the combination of large $\zeta_1$ and small $\zeta_2$ favor the formation of two peaks.
This is not unexpected.
A steeper decline early on accelerates the timescale for light curve evolution (\S\ref{subsec:lc_eeff}), allowing the first peak to fully form, while the much flatter decline at $\tau > \tauq$ provides the additional energy needed to power a second peak. 

While one-zone, gray-opacity models cannot capture the complexities of realistic numerical simulations of transients, like those we present in \S\ref{sec:kilonova}, they are useful for diagnosing trends and understanding important determinants of emission.
We have shown that double peaked light curves are expected analytically when opacity increases in time and the heating rate deviates from a straightforward power-law.
As we argued above, the \rp{}---at least within the limits of current nuclear physics uncertainties---can in some cases meet both of these criteria.
This raises the possibility that kilonova light curves may be much more diverse than previously expected.

\bibliographystyle{apj} 
\bibliography{refs}

\end{document}